\documentclass[3p, review, 12pt]{elsarticle}
\usepackage[utf8]{inputenc}

\title{Image-Histogram-based Secondary Electron Counting to Evaluate Detective Quantum Efficiency in SEM}

\author{Akshay Agarwal}
\author{John Simonaitis}
\author{Karl K. Berggren}
\address{Department of Electrical Engineering and Computer Science, Massachusetts Institute of Technology}

\usepackage{cmap}
\usepackage[utf8]{inputenc}
\usepackage[T1]{fontenc}
\usepackage{graphicx}
\usepackage{siunitx}
\usepackage{cite}
\usepackage{array}
\usepackage{mathtools}
\usepackage{pgfplots}
\pgfplotsset{compat=1.12}
\usetikzlibrary{arrows}
\usepackage{outlines}
\usepackage{enumitem}
\usepackage{matlab-prettifier}
\usepackage[nottoc]{tocbibind}
\usepackage{setspace}
\usepackage{amssymb}
\usepackage{braket}
\usepackage{dcolumn}
\usepackage[titletoc]{appendix}
\usepackage{float}
\usepackage{bm}
\usepackage{caption}
\usepackage{xr}
\makeatletter
\newcommand*{\addFileDependency}[1]{
  \typeout{(#1)}
  \@addtofilelist{#1}
  \IfFileExists{#1}{}{\typeout{No file #1.}}
}
\makeatother

\newcommand*{\myexternaldocument}[1]{%
    \externaldocument{#1}%
    \addFileDependency{#1.tex}%
    \addFileDependency{#1.aux}%
}
\myexternaldocument{supp_info}
 \usepackage[stable]{footmisc}
\def\all{all}
\ifx\files\all  \else 
\def\um{\si{\micro\meter}}

\def\us{\si{\micro\second}}
\def\ns{\si{\nano\second}}

\def\pA{\si{\pico\ampere}}

\def\keV{\si{\kilo\electronvolt}}

\def\NSE{N_{\textrm{SE}}}

\def\SEo{\mathrm{SE_{1}}}
\def\SEt{\mathrm{SE_{2}}}
\def\SEth{\mathrm{SE_{3}}}

\begin{document}



\begin{abstract}
Scanning electron microscopy is a powerful tool for nanoscale imaging of organic and inorganic materials. An important metric for characterizing the limits of performance of these microscopes is the Detective Quantum Efficiency (DQE), which measures the fraction of emitted secondary electrons (SEs) that are detected by the SE detector. However, common techniques for measuring DQE approximate the SE emission process to be Poisson distributed, which can lead to incorrect DQE values. In this paper, we introduce a technique for measuring DQE in which we directly count the mean number of secondary electrons detected from a sample using image histograms. This technique does not assume Poisson distribution of SEs and makes it possible to accurately measure DQE for a wider range of imaging conditions. As a demonstration of our technique, we map the variation of DQE as a function of working distance in the microscope.
\end{abstract}
\maketitle

\section{Introduction}
Scanning electron microscopy (SEM) is a robust and versatile imaging technique that is routinely used to image inorganic and organic samples at resolutions down to 1 nanometer~\citep{Reimer1998e}. In an SEM, a focused probe of electrons at energies typically between 1 and 30 \keV~is raster-scanned across the sample. At each scan position, the incident electron beam generates secondary electrons (SEs), which escape from the sample surface and are detected by SE detectors. 

Efficient detection of SEs is a crucial requirement for full utilization of the high-resolution capabilities of the SEM. This efficiency is measured by a parameter called the Detective Quantum Efficiency (DQE), and it plays a critical role in the signal-to-noise-ratio (SNR) and contrast observed in an SEM image. For low-noise SE detectors, the DQE is defined as the fraction of SEs emitted by a sample pixel that are detected, and it takes values between 0 and 1. Following early work characterizing the efficiency of SE detectors~\citep{Pawley1974, Comins1978, Oatley1981, Oatley1985}, Joy and co-workers~\citep{Joy1996,Joy2007} introduced a technique based on the image pixel brightness histogram for measuring the DQE of SE detectors. Here, the image pixel brightness histogram is a plot of the frequency with which different pixel brightnesses occur in the image (we present an example image histogram of a typical SEM image in Section~\ref{sec:imagehistrobasic} of the Supplementary Information). This work established that variations in the value of DQE between SE detectors are caused by either the type of detection mechanism used or the detector geometry and placement. Therefore, quantitative knowledge of the DQE is important to maximize the performance of SE detectors and troubleshoot poor imaging results in an SEM.

\begin{figure*}[ht]
    \centering
    \includegraphics[scale=0.48]{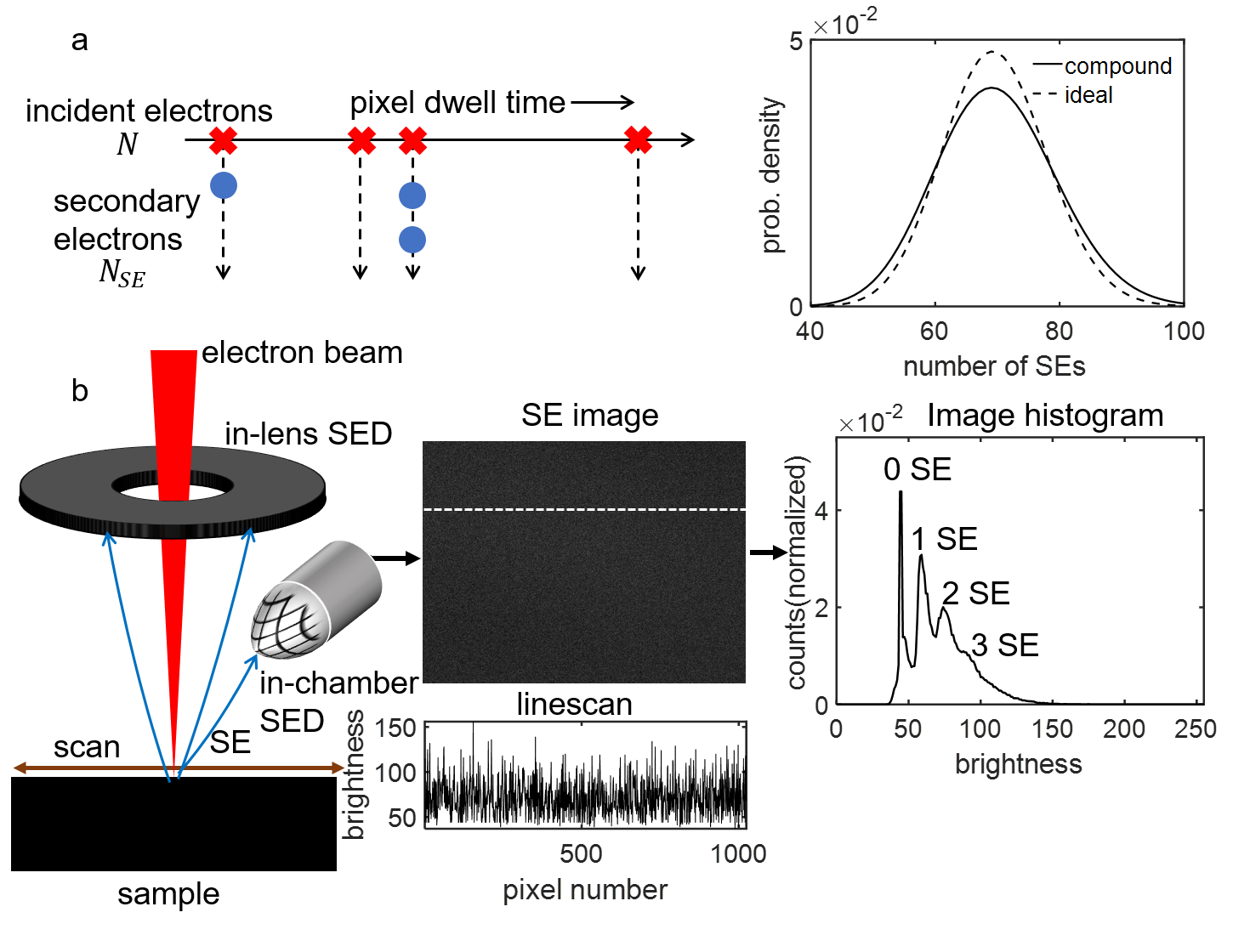}
    \caption[Statistics of secondary electron emission in an SEM]{Statistics of secondary electron emission on an SEM. (a) SE emission is the result of two coupled Poisson processes - the arrival of incident electrons during the pixel dwell time (red crosses) characterized by mean $N$, and the emission of SEs due to the incident electrons (blue circles) characterized by mean $\NSE$. The plot shows a simulation of the resulting compound Poisson distribution of the number of SEs, which has a larger variance than an ideal Poisson distribution with the same mean. This deviation from ideal Poisson distribution needs to be accounted for when measuring SE detector DQE. (b) Experimental setup for measuring DQE. The incident beam scans a featureless sample of aluminum and produces an SE image on both the in-chamber and in-lens detectors. A linescan across the SE image (along the white dashed line) shows variations in the image pixel brightness. These images are used to produce histograms of the pixel brightness. These histograms show discrete peaks corresponding to integral number of SEs, which can be used to directly find $\NSE$ and hence the detector DQE without assuming Poisson statistics.}
    \label{fig:fig1}
\end{figure*}

An important approximation in Joy's technique is that the number of SEs emitted by the sample follows a Poisson distribution. This approximation makes the theoretically expected image SNR proportional to the square root of the expected number of SEs emitted by the sample. Therefore, the DQE calculation is simplified, since DQE can be expressed as the ratio of the experimental SNR (extracted from the peak brightness and width of the image histogram) and the theoretical SNR. However, previous experimental and theoretical work has established that the SE distribution can show significant deviations from Poisson statistics~\citep{Kurrelmeyer1937, Everhart1959, Oatley1981,Oatley1985, Baumann1980,Baumann1981, Novak2009, Frank2005, Sakakibara2019}. These deviations arise because the emission of SEs is the result of two successive Poisson processes, as shown in Figure~\ref{fig:fig1}(a): first, the generation of the incident beam at the electron gun (characterized by the beam current, $I_B$), and second, the emission of SEs by each incident-beam electron (characterized by the SE emission yield, $\delta$). The result of two coupled Poisson processes (also referred to as a compound Poisson process) is not a Poisson process since its variance is higher than its mean. The plot in Figure~\ref{fig:fig1}(a) shows the deviation of a simulated compound Poisson distribution of SEs (emitted at an incident beam current of 70 \pA, pixel dwell time of 440 \ns, SE yield $\delta$ of 0.4, and DQE of 0.9) from an ideal Poisson distribution with the same parameters. We can see that the compound Poisson distribution shows a higher width (\textit{i.e.}, variance) than the ideal Poisson distribution. These deviations increase at low ($<5$ \keV) incident beam energies due to higher SE yield $\delta$~\citep{Frank2005,Seiler1983}.

Here, we will present an alternative method of measuring the DQE of SE detectors which does not use the assumption of Poisson statistics. In this method, we will directly count the mean number of SEs emitted from the sample by optimizing imaging conditions to observe peaks in the histograms of SE images due to integral numbers of SEs. Our method of measuring DQE combines and extends observations by Joy and co-workers~\citep{Joy2007} and facilitates the measurement of DQE for a wide range of beam energies and currents on the SEM. As a demonstration of our technique, we will compare the DQE of the in-chamber and in-lens SE detectors on an SEM as a function of the sample working distance.

\section{Methods}
\label{sec:methods}
To motivate our experimental method, we will first consider the emission and detection of SEs in an SEM analytically. There are two sources of SEs detected on the SE detectors. First, as discussed earlier, the incident electron beam generates SEs from the sample surface. These SEs are referred to as $\SEo$s. Second, SEs are also generated by some of the incident electrons which undergo multiple scattering events inside the sample and re-emerge from the sample surface as back-scattered electrons (BSEs). BSEs can lead to SE emission from the sample ($\SEt$s) as well as the SEM chamber walls and objective lens polepiece ($\SEth$s). Detectors for SEs are typically placed either inside the vacuum chamber (the \textit{in-chamber} SE detector) or inside the polepiece of the objective lens (the \textit{in-lens} SE detector). Images generated by the in-lens SE detector are usually higher-resolution because this detector preferentially detects $\SEo$s and $\SEt$s inside the sample, and filters out most of the lower-resolution $\SEth$s due to its position inside the lens polepiece~\citep{Griffin2011}. 

The mean number of incident electrons $N$ on a sample pixel is given by:
\begin{equation*}
    N=\frac{I_B\tau}{e}.
\end{equation*}
Here, $\tau$ is the pixel dwell time, and $e =1.602 \times 10^{-19}$ C is the electron charge. Then, the average number of SEs from an object pixel, $N_{SE,\textrm{object}}$, is given by:
\begin{equation*}
    N_{SE,\textrm{object}}=N\cdot\delta=\frac{I_B\tau}{e}\cdot\delta.
\end{equation*}
Here $\delta$ is the total SE yield of the object pixel. Finally, the average number of SEs detected by the SE detector from that pixel, $N_{SE}$, is given by:
\begin{equation}
    N_{SE}=N_{SE,\textrm{object}}\cdot\textrm{DQE}=\frac{I_B\tau}{e}\cdot\delta\cdot\textrm{DQE}.
    \label{eqn:DQE}
\end{equation}

Therefore, the aim of our experiments was to extract DQE by measuring $N_{SE}$ at different values of $I_B$ for a sample with known $\delta$.

Figure~\ref{fig:fig1}(b) shows our experimental setup. We scanned a featureless, bulk piece of aluminum at an incident beam energy of 10 \keV~on a Zeiss LEO 1525 SEM equipped with a Schottky electron gun. An aluminum sample was used because the SE yield of aluminum is well-characterized (we used $\delta=0.2$ in our analysis~\citep{Seiler1983}), and the lack of features on the sample ensured that the SE yield was constant across the entire scan region on sample. Further, we scanned the sample at low magnification (pixel size $>1$~\um) so that the beam was sufficiently defocused to smooth out local fluctuations in sample $\delta$. We used either the in-chamber or the in-lens SE detector to generate an SE image of the featureless sample. The sample image in Figure~\ref{fig:fig1}(b) appears smooth due to the lack of features in the sample; a linescan across the image reveals variations in the pixel brightness. The SEM in our experiments generated 8-bit, $1024\times768$ images, and therefore the range of pixel brightness values was 0 to 255. We obtained a histogram of pixel brightness from such SE images. As shown in the example in Figure~\ref{fig:fig1}(b), the image histogram showed discrete peaks due to integral number of SEs, which we used to calculate the mean number of SEs emitted by the sample and hence the DQE. 

The observation of quantized SE peaks was central to our method of calculating DQE. Although such peaks were reported previously by Joy~\citep{Joy2007}, they were not well-resolved and could not be used for quantitative analysis of the statistics of SE emission.
To reliably observe and resolve these SE peaks, we ensured that $N_{SE}$ was low by lowering the incident beam current $I_B$ and/or the pixel dwell time $\tau$, in accordance with Equation~\ref{eqn:DQE}. To lower $I_B$, we used the smallest current-limiting aperture (with a diameter of 7.5 \um) on the SEM. We optimized $I_B$ by changing the gun extraction voltage, and the pixel dwell time $\tau$ by changing the scan speed, to obtain integral-SE-number peaks in the image histogram. Further, we chose an incident beam energy of 10 kV to ensure that $\delta$ was low enough to resolve single SEs. We also optimized the imaging brightness and contrast, as described in more detail in Section~\ref{sec:imageopt} of the Supplementary Information.

\begin{figure*}[ht]
    \centering
    \includegraphics[scale=0.47]{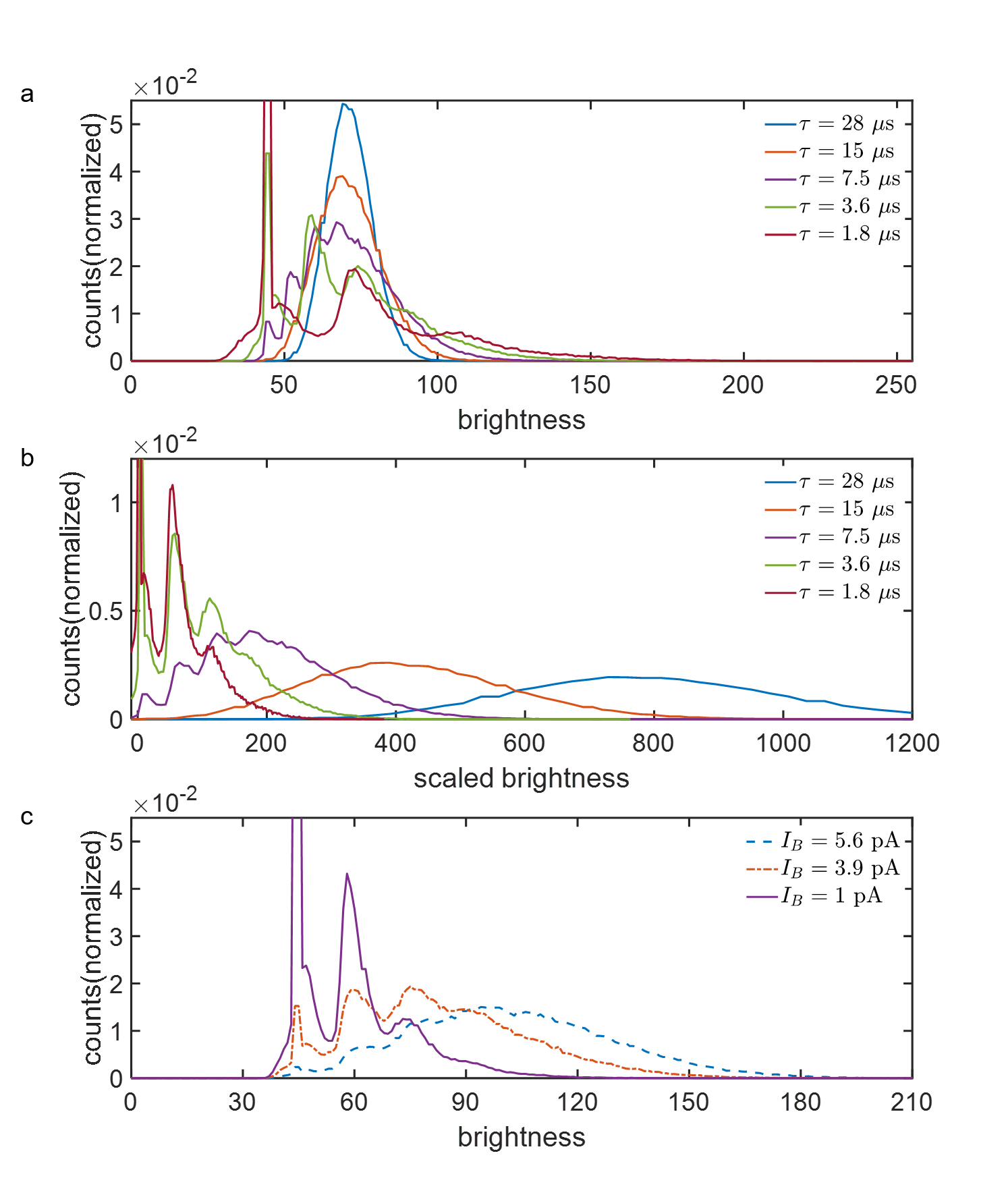}
    \caption[SE quantization observed on the in-chamber detector in SEM]{SE quantization observed on the in-chamber detector in SEM. (a) Variation of the image histogram with pixel dwell time $\tau$. As we reduced $\tau$, distinct peaks appeared in the image histogram corresponding to integral numbers of detected SEs. (b) Histograms from (a) with noise peak at 44 shifted to zero, and scaled by the pixel dwell time. The integral SE number peaks are aligned with each other. (c) Variation of the image histogram with incident beam current $I_B$ for constant $\tau$. As $I_B$ reduced, the SE peaks became sharper and more well-resolved, indicating that the sample emitted fewer SEs.} 
    \label{fig:fig2}
\end{figure*}

Figure~\ref{fig:fig2}(a) shows image histograms for different values of $\tau$, made using SE images collected on the in-chamber detector. This experiment was carried out at $I_B \approx 2.3$ \si{\pico\ampere}. The different histograms correspond to a $\tau$ of \num{28} \si{\micro\second} (blue curve), \num{15} \si{\micro\second} (orange curve), \num{7.5} \si{\micro\second} (purple curve), \num{3.6} \si{\micro\second} (green curve), and \num{1.8} \si{\micro\second} (dark red curve). Note that each histogram is normalized so that the total area under the histogram is 1. The histogram for $\tau = 28$ \us~shows a single peak at a pixel brightness of 68. As we lowered $\tau$ from \num{28} to \num{15} \si{\micro\second} the histogram became wider but retained the same mean. At $\tau=$\num{7.5} \si{\micro\second}, a series of sharp peaks appeared in the histogram. The first of these peaks was centered at a brightness level of \num{44}, the second at \num{51}, and the third at \num{58}. Upon reducing $\tau$ to \num{3.6} \si{\micro\second}, the first peak remained at \num{44}, but the second shifted to \num{58}, and the third to \num{73}. This shift continued upon further reduction of the dwell time to \num{1.8} \si{\micro\second}, in which case the second peak appeared at \num{73} and the third at \num{105}. Simultaneously, as $\tau$ was reduced, the first peak at brightness \num{44} increased in intensity.

We attributed the peaks that emerged in the histogram at small dwell times to integral numbers of SEs. The constant sharp peak at pixel brightness \num{44} can be attributed to noise and corresponds to pixels with zero detected SEs, since all other peaks disappeared when the beam was turned off (see Section~\ref{subsec:blankingop} for more details). The shifting of the other peaks' pixel brightness levels when the dwell time was lowered can be explained as a consequence of signal time-averaging, which is a correction of the observed signal level to account for varying pixel dwell time performed on many electron microscopes~\citep{Sang2016}.

In Figure~\ref{fig:fig2}(b), we re-plot each of the image histograms from Figure~\ref{fig:fig2}(a), multiplied by the corresponding pixel dwell time, with the zero-SE peak shifted to 0. The horizontal axis corresponds to the new, scaled values of brightness. The histograms were also re-normalized with respect to this scaled brightness. Since this re-scaling removes the effect of signal-time averaging, the reduction in mean scaled brightness (caused by reduction in signal level) with reduction in pixel dwell times can now be seen. We also see that the integral SE number peaks for $\tau=$ \num{7.5}~\us, \num{3.6}~\us~and \num{1.8}~\us~are now at the same scaled brightness values, and the gaps between the peaks are now the same for all dwell times.

In Figure~\ref{fig:fig2}(c), we show the variation of the histogram with incident beam current $I_B$ for the in-chamber detector. As mentioned previously, we controlled $I_B$ by changing the gun extraction voltage. These measurements were made at $\tau=$ \num{3.6} \si{\micro\second}. The values of $I_B$ for which we plotted histograms in this figure are \num{5.6} \si{\pico\ampere} (dashed blue curve), \num{2.9} \si{\pico\ampere} (dash-dotted orange curve), and \num{1} \si{\pico\ampere} (solid purple curve). We can see that as $I_B$ reduces, the one-SE peak at pixel brightness \num{58} becomes stronger, while the peaks representing three and more SE peaks become weaker. The two-SE peak is stronger for \num{2.9} \si{\pico\ampere} than the other incident beam current values. These trends indicate that the mean number of detected SEs reduced upon reducing $I_B$. In Section~\ref{sec:results} we will analyze this reduction quantitatively and see that it is linear at low incident beam currents, as we would expect from Equation~\ref{eqn:DQE}. We obtained similar results for the in-lens detector, as described in Section~\ref{sec:inlens} of the Supplementary Information.

Our observations from Figure~\ref{fig:fig2} support the conclusion that the peaks in the histogram are due to single SEs. To further support this conclusion, we coupled the signal from the in-chamber SE detector onto an oscilloscope and created histograms of the average signal level observed on the histogram for a time window corresponding to the pixel dwell time. We observed that these histograms closely matched the image histograms obtained from the same scan areas. These experiments are described in greater detail in Section~\ref{sec:oscilloscope} of the Supplementary Information.

Based on the results in Figure~\ref{fig:fig2}, we decided to use  $\tau=$ \num{3.6} \si{\micro\second} as the pixel dwell time for future experiments since this dwell time gave us strong multi-electron peaks that were well-resolved. We varied $I_B$ depending on the desired mean number of SEs per pixel by tuning the gun extraction voltage.

\section{Results}
\label{sec:results}
In the previous section, we showed how quantized SE peaks emerged in the image histogram when $I_B$ and $\tau$ were optimized. In this section, we will use the image histogram peaks to analyze the statistics of the SEs. We will look at the mean SE number $N_{SE}$ obtained from the image histograms and use it to calculate the DQE of the in-chamber and in-lens detectors. We will also compare the DQE obtained for these detectors with values obtained using Joy's method~\citep{Joy1996,Joy2007}. Finally, we will look at the variation of the DQE with sample working distance.

\subsection{SE counting for in-chamber detector image histograms}
Figure~\ref{fig:fig3}(a) is an image histogram for the in-chamber detector at an incident beam energy of 10 keV, $I_B=2.3$ \pA~and $\tau=3.6$ \us. As mentioned before, the histogram is normalized so that the total area under it is 1. Therefore, we can treat this histogram as a probability mass function and find the mean pixel brightness. Knowing the zero-, one-, and two-SE levels, we can translate pixel brightness values to a corresponding SE number, as indicated on the horizontal axis of Figure~\ref{fig:fig3}(a). Using this new horizontal axis, we can find the mean SE number $\NSE$. For the histogram in Figure~\ref{fig:fig3}(a), this mean SE number is \num{1.91}, indicated by the black vertical line. 

In Figure~\ref{fig:fig3}(b), we plot the mean SE number extracted from image histograms for a range of incident beam currents, for both the in-chamber detector (unfilled black circles) and the in-lens detector (red crosses). The data point corresponding to the histogram in Figure~\ref{fig:fig3}(a) is indicated with a filled black circle. As we lowered the beam current from 7.8 \si{\pico\ampere} to 0.5 \si{\pico\ampere}, the mean SE number reduced from 5.31 to 0.48 for the in-chamber detector and from 9.83 to 1.04 for the in-lens detector. We see that the mean SE number varies linearly with $I_B$ for low currents and shows some non-linearity at higher currents for both detectors. The solid black line is a least-square fit to the extracted SE number for $I_B<5$ \pA~(indicated by the vertical dotted black line) for the in-chamber detector, and the solid red line is a fit to the extracted SE number for the same range of incident beam current for the in-lens detector. We expect that the non-linearity at higher currents is caused by increasing incidents of multiple SEs hitting the detector within a fraction of the pixel dwell time, causing signal pileup and consequent incorrect extracted SE values. We discuss the mechanism for this pileup in more detail in Section~\ref{sec:oscilloscope} of the Supplementary Information. In this paper, we restricted ourselves to a very small probability of multiple incidence by using $I_B<5$ \pA~for quantitative analysis of DQE.

We note that the mean SE number is higher by a factor of 2 for the in-lens detector than for the in-chamber detector. The higher mean SE number indicates that the in-lens detector is more efficient at collecting SEs than the in-chamber detector, as reported previously~\citep{Joy1996, Griffin2011}. 

\subsection{DQE from in-chamber and in-lens detector image histograms}
\label{subsubsec:imagehistodqe}

\begin{figure*}
    \centering
    \includegraphics[scale=0.48]{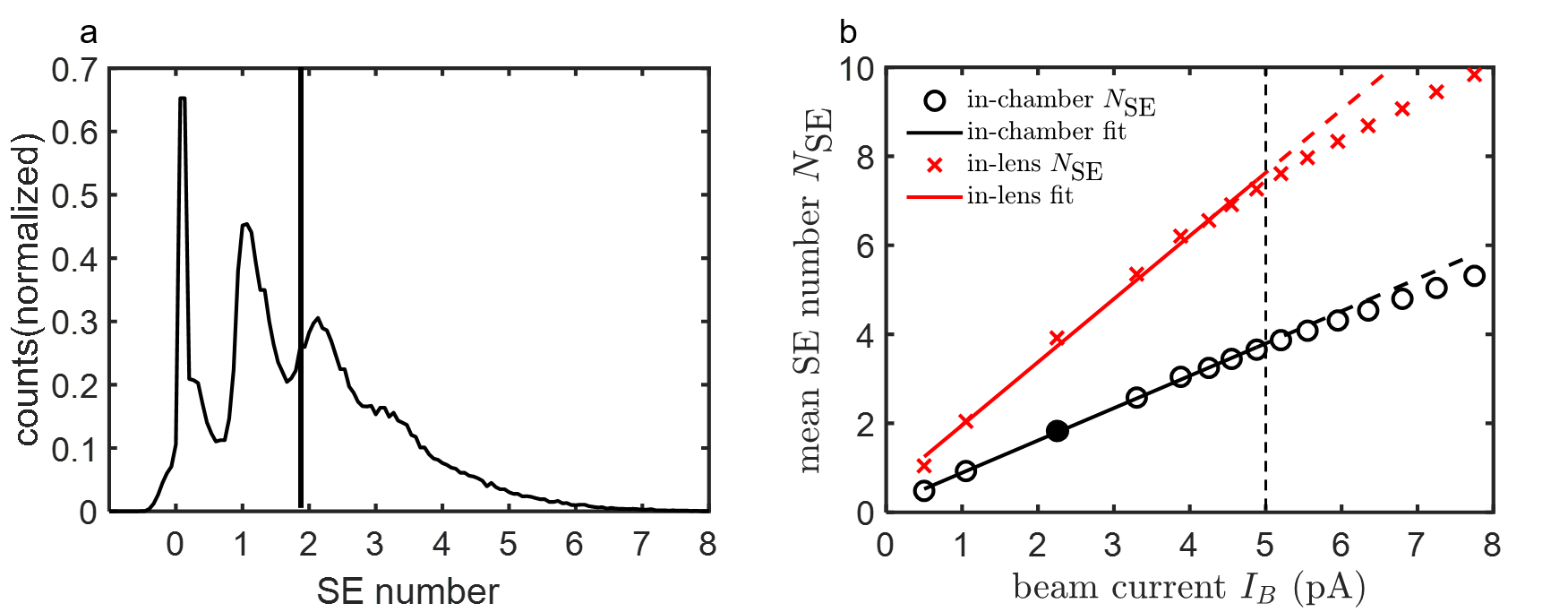}
    \caption[Statistics of image histogram SE counting]{Statistics of image histogram SE counting. (a) Example image histogram for the in-chamber detector, with a mean SE number of 1.91 (indicated by the vertical black line). (b) Variation of mean SE number with incident beam current for the in-lens and in-chamber detectors. Unfilled black circles indicate extracted $\NSE$ values for the in-chamber detector, and red crosses indicate $\NSE$ values for the in-lens detector. Filled black circle indicates $\NSE$ for the histogram in (a). The least-squares fit lines (black for the in-chamber detector and red for the in-lens detector) are for $I_B<5 $ \si{\pico\ampere}, indicated by the vertical dashed black line. Above this incident beam current, the extracted $\NSE$ values deviate from this fit due to signal pileup. The DQE extracted from the slope of these lines is 0.16 for the in-chamber detector and 0.32 for the in-lens detector.}
    \label{fig:fig3}
\end{figure*}

We can use the least-square fits for the in-chamber and in-lens detectors in Figure~\ref{fig:fig3}(b) to extract the DQE for these detectors. From Equation~\eqref{eqn:DQE}, the mean SE number $N_{SE}$ is given by:

\begin{equation*}
    N_{SE}=\frac{I_B\tau}{e}\cdot\delta\cdot\textrm{DQE}.
\end{equation*}

As mentioned before, we used $\tau=3.6$ \si{\micro\second} for all the image histograms used to extract the data shown in Figure~\ref{fig:fig3}. We used a value of 0.2 for the SE yield $\delta$ of our aluminum sample at 10 kV~\citep{Seiler1983, Frank2005, Reimer1998}. We used these values to extract the DQE from the slope of the least-squares-fit lines in Figure~\ref{fig:fig3}(b). The extracted values were 0.16 for the in-chamber detector and 0.32 for the in-lens detector. In order to benchmark these DQE values, we implemented Joy's DQE method, as detailed in Section~\ref{sec:joymethod} of the Supplementary Information. The DQE values obtained from our method are close to the range of values obtained using Joy's method (between 0.15 and 0.22 for the in-chamber detector and between 0.3 and 0.6 for the in-lens detector), and are also in the range of reported values in the literature for these detectors~\citep{Joy1996, Joy2007}. As discussed by Joy, DQE values can vary by orders of magnitude depending on the detector geometry and age~\citep{Joy2007}. As an example of another source of this variation, in the next section we will see how the DQE of both detectors varied with the working distance.

\subsection{Variation of DQE with working distances}
\begin{figure*}[ht]
    \centering
    \includegraphics[scale=0.48]{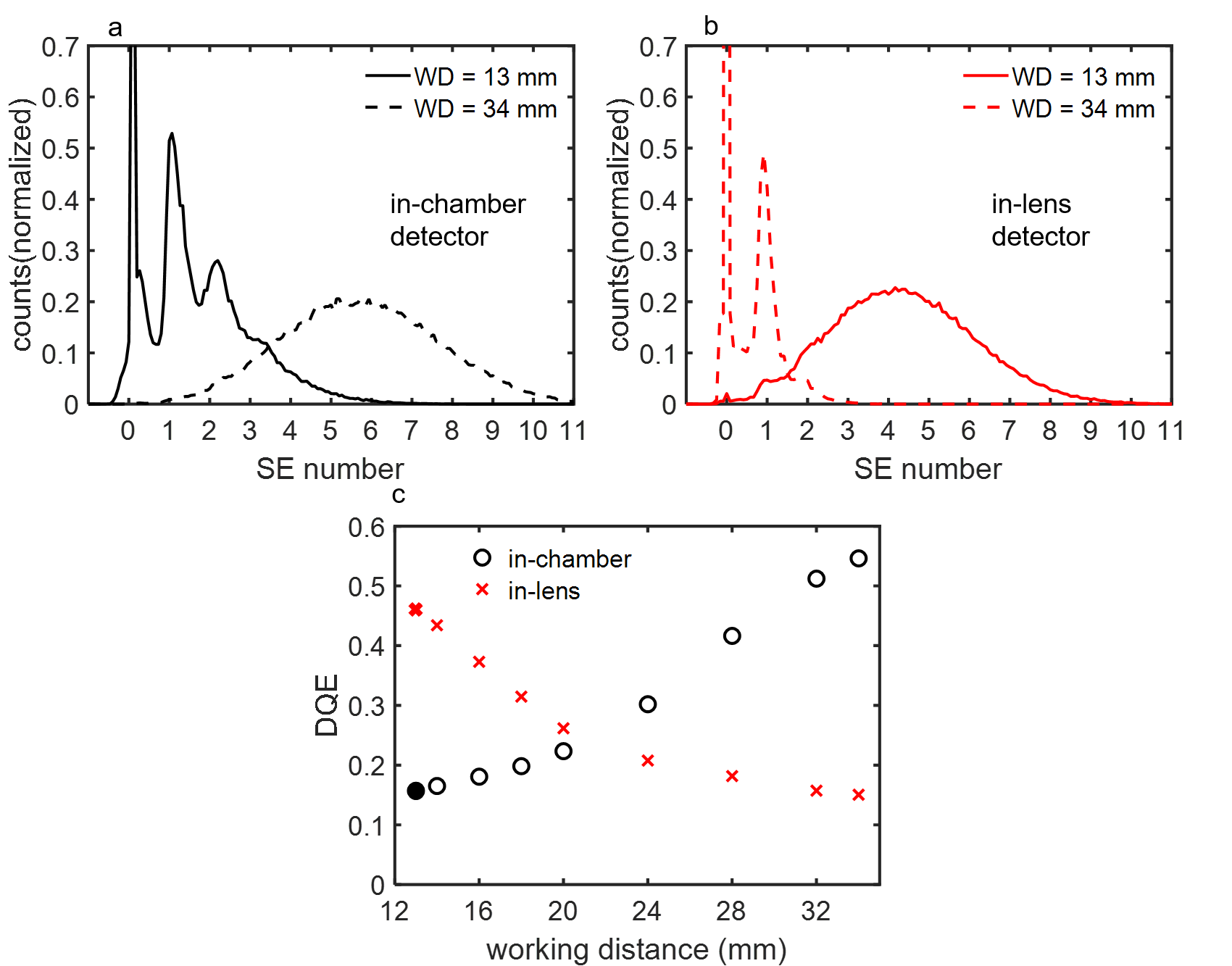}
    \caption[Variation of DQE with working distance for in-chamber and in-lens SE detectors]{Variation of DQE with working distance for in-chamber and in-lens SE detectors. (a) Image histograms for SEM images captured using the in-chamber detector for working distances of 13 mm (solid black curve) and 34 mm (dash-dotted black curve). The mean SE number is higher at 34 mm than at 13 mm. (b) Image histograms for SEM images captured using the in-lens detector for working distances of 13 mm (solid red curve) and 34 mm (dash-dotted red curve). The mean SE number is lower at 34 mm than at 13 mm. (c) DQE for the in-chamber detector (unfilled black circles) and in-lens detector (red crosses) at different working distances. The DQE for the in-chamber detector increases with working distance, while the DQE for the in-lens detector goes down. The filled black circle and bold red cross indicate DQE for the working distance of 13 mm, which was used to obtain all the results in previous figures.}
    \label{fig:fig4}
\end{figure*}
As an application of this technique, we investigated the variation of the in-chamber and in-lens detector DQE as a function of the working distance. The SEM we used has an 8 kV electrostatic field along the optical axis to attract SEs to the in-lens detector. Working with a similar column design, Griffin~\citep{Griffin2011} has previously reported qualitative improvement in the fine details observable in the in-chamber detector image with increasing working distance. Through our measurement of the DQE for both detectors, we wanted to quantify these changes in the detector efficiency as a function of working distance.

In Figure~\ref{fig:fig4}(a) we plot the image histograms for the in-chamber detector at two working distances (WD): 13 mm (solid black curve) and 34 mm (dash-dotted black curve), at a beam current of 2.3 \pA.  We see that the SE peaks are well-defined for the lower working distances but not the higher working distance. This observation indicates that the mean SE number was higher at $\textrm{WD} = 34$ mm than at $\textrm{WD} = 13$ mm. From the histograms, $N_{SE}=5.78$ for WD $=34$ mm and $N_{SE}=1.66$ for WD $=13$ mm. This increase in mean SE number (and consequent increase in DQE) translates to higher SNR, consistent with Griffin's observation. 

In Figure~\ref{fig:fig4}(b), we plot the image histograms at $\textrm{WD} = 13$ mm (solid red curve) and $\textrm{WD} = 34$ mm (dash-dotted red curve) for the in-lens detector. Contrary to our observations with the in-chamber detector, for the in-lens detector $N_{SE}$ was lower at $\textrm{WD} = 34$ mm, where it is 0.32, than at $\textrm{WD} = 13$ mm, where it was 4.42. This effect was not reported by Griffin, who noted little variation in the in-lens image with increasing working distance. 

Figure~\ref{fig:fig4}(c) is a plot of the DQE for both detectors for working distances between 13 mm and 34 mm. The DQE values calculated from the image histograms in Figure~\ref{fig:fig3}, which were obtained at a working distance of 13 mm, are indicated by the filled black circle and the bold red cross. We can see the trend of increasing DQE for the in-chamber detector and decreasing DQE for the in-lens detector with increasing working distance. We believe that the reduction in DQE for the in-lens detector with increasing working distance is a result of the diminishing effect of the 8 kV electrostatic field designed to attract SEs to the in-lens detector. At higher working distances, this field would be weaker and less effective at drawing SEs back up the column to the in-lens detector. The weaker field would also result in more electrons being available to be detected by the in-chamber detector which explains part of the increase in its DQE with increasing working distance. A second reason for the increase in the DQE of the in-chamber detector with working distance could be that at larger working distances, the detector geometry and location with respect to the sample becomes more favorable and allows more SEs to reach the detector. Further, as the working distance increases, the solid angle within which most of the BSEs are emitted covers more of the surface of the objective lens polepiece and chamber walls of the SEM. Therefore, the BSEs can generate more $\SEth$s which would be detected by the in-chamber detector. The in-lens detector, placed along the optical axis so that it is much less sensitive to $\SEth$s by design, would not see this increase. We believe that the reduction in in-lens detector DQE (and consequent reduction in SNR) was not seen by Griffin because of higher incident beam currents which ensured that even at large working distances, sufficient $\SEo$s were detected at the in-lens detector to generate a high-SNR image.

\section{Conclusions}
In this paper, we have shown how image histogram-based SE counting can be used to calculate the DQE of SE detectors in an SEM. Our DQE calculation method builds on work by Joy et al.~\citep{Joy1996, Joy2007}, Timischl et al.~\citep{Timischl2012}, and Griffin~\citep{Griffin2011}. We also characterized the variation of the DQE with working distance for both in-chamber and in-lens SE detectors. Our analysis of the variation of the detector DQE with working distance revealed that the DQE for the in-chamber detector increases and the DQE for the in-lens detector falls with increasing working distance. We speculated that the decrease in the in-lens DQE arose from the reduced effectiveness of the 8 kV in-lens detector attraction field at higher working distances, and the increase in the in-chamber detector DQE arose from greater availability of SEs, better geometry, as well as greater generation of $\SEth$s by BSEs. Establishing which of these effects is dominant by, for example, varying the in-lens detector field would be a useful extension of this work. 

Unlike the work of Joy and co-workers, we did not use image histograms to find an SNR in our work. We found that the widths (\textit{i.e.,} variances) of the probability mas functions extracted from the image histograms were an order of magnitude higher than the mean SE counts. We were unsure of the reason for these large image histogram widths. Due to these large variances we could not extract reliable SNR values from the image histograms. A more detailed analysis of higher-order statistics of the image histograms could be the subject of future work.

The appearance of quantized SE peaks in the image histogram can be used to create SE count images on the SEM. SE count imaging has been suggested as a high-SNR extension of conventional SE imaging~\citep{Joy2007}, and was implemented by Yamada and co-workers~\citep{Yamada1990,Yamada1990a,Yamada1991,Yamada1991a,Uchikawa1992} using external pulse-counting and timing circuits synchronized with the SEM. A histogram-based SE counting method would be direct, require no external circuitry, and could be incorporated into commercial SEM software. Such a technique would generate SE count images by assigning an SE number to each pixel in the image based on its pixel brightness, knowing the positions of the integral-number-SE peaks. However, as can be seen from the image histograms in Figures~\ref{fig:fig2},~\ref{fig:fig3}, and~\ref{fig:fig4}, there is significant overlap between the one-, two-, and three-SE peaks, and assigning an SE number to pixels with brightnesses intermediate between the discrete peak regions would require a probabilistic decision scheme. Such a scheme would lead to errors in the assignment of some pixels. However, if a few errors are acceptable as a trade-off for ease of implementation, a histogram SE count imaging scheme is an attractive option for extending conventional SEM imaging. In our work, the single SE signal was observable in the image histogram for pixel dwell times down to 1 \si{\micro\second}. Although we did not use incident beam currents as low as in Yamada's work (the lowest current we used was 0.55 \si{\pico\ampere} compared to 0.1 \si{\pico\ampere} by Yamada), combining small pixel dwell times with low incident beam currents could lead to live imaging of radiation-sensitive samples such as proteins and biomolecules at extremely low incident electron doses and dose rates. Such low dose-rate imaging could also be applied to time-resolved SEM and ion-beam imaging~\citep{Medin2018a,Medin2019,Peng2020SourceMeasurement}.


\section*{Acknowledgements}
The authors thank Navid Abedzadeh for help with initial SEM histogram measurements, and the QEM-2 collaboration for insightful discussions. The authors also acknowledge Marco Turchetti, Yujia Yang, Vivek K. Goyal and James LeBeau for helpful feedback. This work was supported by the Gordon and Betty Moore Foundation. This material is based upon work supported by the National Science Foundation Graduate Research Fellowship under Grant No. 1745302.

\bibliography{references}

\begin{thebibliography}{10}

\bibitem{Reimer1998e}
Ludwig Reimer.
\newblock {Introduction}.
\newblock In {\em Scanning Electron Microscopy}, pages 1--12. Springer-Verlag
  Berlin Heidelberg New York, 2nd edition, 1998.

\bibitem{Pawley1974}
J.~B. Pawley.
\newblock {Performance of SEM Scintillation Materials}.
\newblock In {\em Scanning Electron Microscopy 1974 (Part 1) - Proceedings of
  the Seventh Annual Scanning Electron Microscopy Symposium, IIT Research
  Institute, Chicago, Illinois 60616, USA}, pages 27--34, 1974.

\bibitem{Comins1978}
N.~R. Comins, M.~M.~E. Hengstberger, and J.~T. Thirlwall.
\newblock {Preparation and evaluation of P-47 scintillators for a scanning
  electron microscope}.
\newblock {\em Journal of Physics E: Scientific Instruments},
  11(10):1041--1047, 1978.

\bibitem{Oatley1981}
C.~W. Oatley.
\newblock {Detectors for the scanning electron microscope}.
\newblock {\em Journal of Physics E: Scientific Instruments}, 14:971--976,
  1981.

\bibitem{Oatley1985}
C.~W. Oatley.
\newblock {The detective quantum efficiency of the scintillator/
  photomultiplier in the scanning electron microscope}.
\newblock {\em Journal of Microscopy}, 139(2):153--166, 1985.

\bibitem{Joy1996}
D~C Joy, C~S Joy, and R~D Bunn.
\newblock {Measuring the performance of scanning electron microscope
  detectors}.
\newblock {\em Scanning}, 18(8):533--8, 1996.

\bibitem{Joy2007}
David~C. Joy.
\newblock {Noise and Its Effects on the Low-Voltage SEM}.
\newblock In {\em Biological Low-Voltage Scanning Electron Microscopy}, pages
  129--144. Springer New York, New York, NY, 2007.

\bibitem{Kurrelmeyer1937}
Bernhard Kurrelmeyer and Lucy~J Hayner.
\newblock {Shot Effect of Secondary Electrons from Nickel and Beryllium}.
\newblock {\em Physical Review}, 52(1937):952--958, 1937.

\bibitem{Everhart1959}
T.~E. Everhart, O.~O. Wells, and C.~W. Oatley.
\newblock {Factors affecting contrast and resolution in the scanning electron
  microscope}.
\newblock {\em Journal of Electronics and Control}, 7(2):97--111, 1959.

\bibitem{Baumann1980}
W.~Baumann, A.~Niemietz, L.~Reimer, and B.~Volbert.
\newblock {Noise Measurements of Different SEM Detectors}.
\newblock {\em Electron Microscopy 1980}, 3:174--175, 1980.

\bibitem{Baumann1981}
W.~Baumann and L.~Reimer.
\newblock {Comparison of the Noise of Different Electron Detection Systems
  Using a Scintillator-Photomultiplier Combination}.
\newblock {\em Scanning}, 4:141--151, 1981.

\bibitem{Novak2009}
L.~Novak and I.~M{\"{u}}llerova.
\newblock {Single electron response of the scintillator-light
  guide-photomultiplier detector}.
\newblock {\em Journal of Microscopy}, 233(1):76--83, 2009.

\bibitem{Frank2005}
Luděk Frank.
\newblock {Noise in secondary electron emission : the low yield case}.
\newblock {\em Journal of Electron Microscopy}, 54(4):361--365, 2005.

\bibitem{Sakakibara2019}
Makoto Sakakibara, Makoto Suzuki, Kenji Tanimoto, Yasunari Sohda, Daisuke
  Bizen, and Koji Nakamae.
\newblock {Impact of secondary electron emission noise in SEM}.
\newblock {\em Microscopy}, 68(4):279--288, 2019.

\bibitem{Seiler1983}
H.~Seiler.
\newblock {Secondary electron emission in the scanning electron microscope}.
\newblock {\em Journal of Applied Physics}, 54(11):R1, 1983.

\bibitem{Griffin2011}
Brendan~J. Griffin.
\newblock {A comparison of conventional Everhart-Thornley style and in-lens
  secondary electron detectors-a further variable in scanning electron
  microscopy}.
\newblock {\em Scanning}, 33(3):162--173, 2011.

\bibitem{Sang2016}
Xiahan Sang and James~M Lebeau.
\newblock {Characterizing the response of a scintillator-based detector to
  single electrons}.
\newblock {\em Ultramicroscopy}, 161:3--9, 2016.

\bibitem{Reimer1998}
Ludwig Reimer.
\newblock {Emission of Backscattered and Secondary Electrons}.
\newblock In {\em Scanning Electron Microscopy}, pages 135--169.
  Springer-Verlag Berlin Heidelberg New York, 2nd edition, 1998.

\bibitem{Timischl2012}
F.~Timischl, M.~Date, and S.~Nemoto.
\newblock {A statistical model of signal-noise in scanning electron
  microscopy}.
\newblock {\em Scanning}, 34(3):137--144, 2012.

\bibitem{Yamada1990}
S.~Yamada, T.~Ito, K.~Gouhara, and Y.~Uchikawa.
\newblock {Electron Counting for Secondary Electron Detection in SEM}.
\newblock {\em Scanning}, 12(5):I--28 -- I--29, 1990.

\bibitem{Yamada1990a}
S.~Yamada, T.~Ito, K.~Gouhara, and Y.~Uchikawa.
\newblock {Secondary Electron Counting Images in SEM}.
\newblock In {\em Proceedings of the XIIth International Congress for Electron
  Microscopy}, pages 402--403, 1990.

\bibitem{Yamada1991}
S.~Yamada, T.~Ito, K.~Gouhara, and Y.~Uchikawa.
\newblock {Electron-Count Imaging in SEM}.
\newblock {\em Scanning}, 13:165--171, 1991.

\bibitem{Yamada1991a}
S.~Yamada, T.~Ito, K.~Gouhara, and Y.~Uchikawa.
\newblock {High-Speed Electron Counting System for TV-Scan Rate SE Images of
  SEM}.
\newblock In {\em Proceedings of the 49th Annual Meeting of the Electron
  Microscopy Society of America}, pages 512--513, 1991.

\bibitem{Uchikawa1992}
Yoshiki Uchikawa, Kazutoshi Gouhara, Satoru Yamada, Tsutomu Ito, Tetsuji
  Kodama, and Pooja Sardeshmukh.
\newblock {Comparative Study of Electron Counting and Conventional Analogue
  Detection of Secondary Electrons in SEM}.
\newblock {\em Journal of Electron Microscopy}, 41(4):253--260, 1992.

\bibitem{Medin2018a}
Safa~C. Medin, John Murray-Bruce, and Vivek~K. Goyal.
\newblock {Optimal stopping times for estimating bernoulli parameters with
  applications to active imaging}.
\newblock {\em ICASSP, IEEE International Conference on Acoustics, Speech and
  Signal Processing - Proceedings}, 2018-April(Section 2):4429--4433, 2018.

\bibitem{Medin2019}
Safa~C. Medin, John Murray-Bruce, David Castanon, and Vivek~K Goyal.
\newblock {Beyond Binomial and Negative Binomial: Adaptation in Bernoulli
  Parameter Estimation}.
\newblock {\em IEEE Transactions on Computational Imaging}, 5(4):570--584,
  2019.

\bibitem{Peng2020SourceMeasurement}
Minxu Peng, John Murray-Bruce, Karl~K. Berggren, and Vivek~K Goyal.
\newblock {Source Shot Noise Mitigation in Focused Ion Beam Microscopy by
  Time-Resolved Measurement}.
\newblock {\em Ultramicroscopy}, page 112948, 2020.

\end{thebibliography}
\bibliographystyle{unsrt}
\end{document}


\maketitle

\section{Introduction to histograms of SEM images}
\label{sec:imagehistrobasic}
\begin{figure}
    \centering
    \includegraphics[scale=0.4]{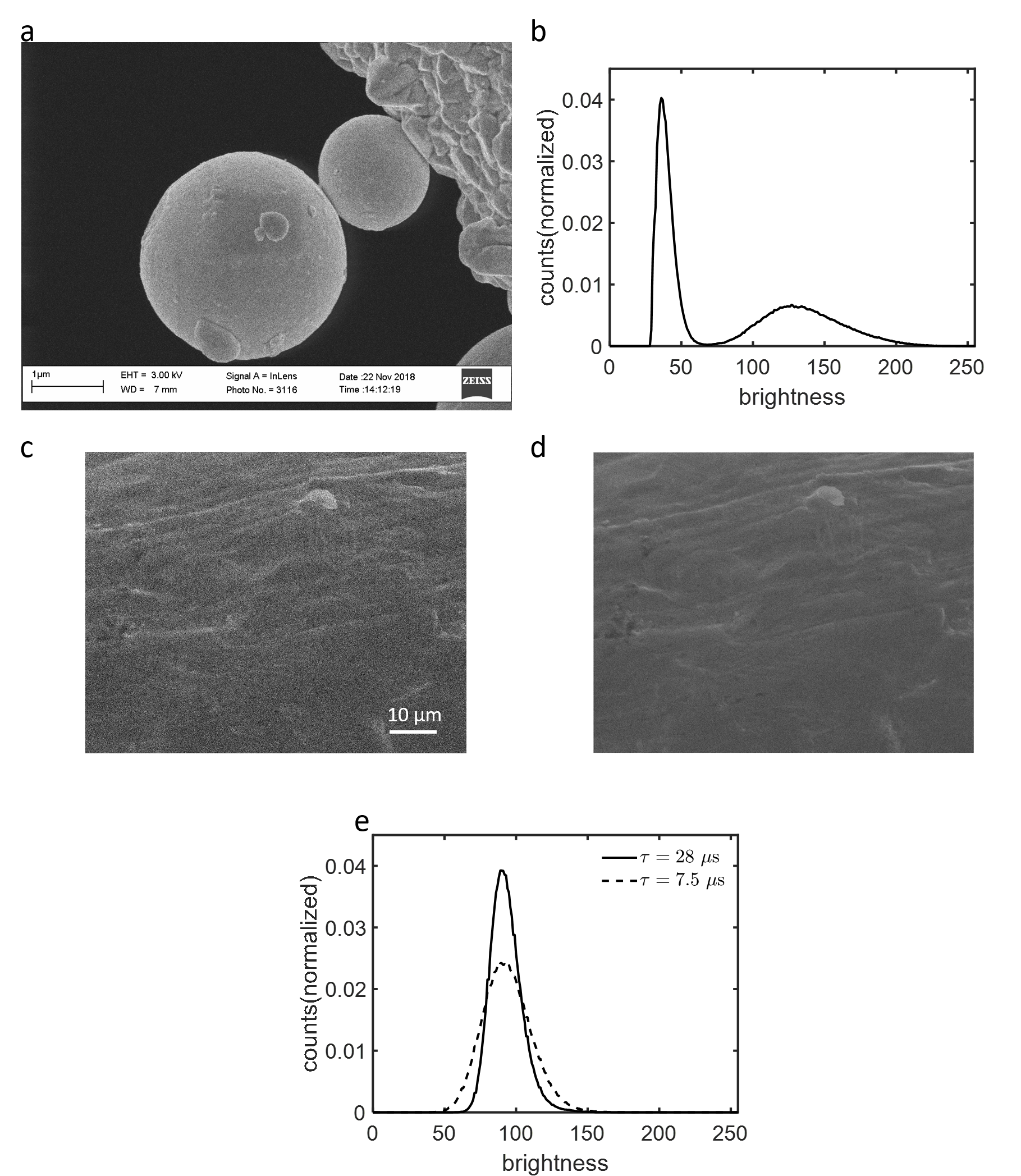}
    \caption[Image histograms in SEM]{Image histograms in SEM. (a) SEM image of spherical tin nanoparticles on a copper TEM support grid. (b) Histogram of the SEM image in (a), showing two peaks due to the background and sample pixels in the image. (c) SEM image of a uniform region of copper taken at $\tau=7.5$ \us. (d) SEM image of the same uniform region of copper as (c) taken at $\tau=28$ \us. The image is less noisy than (c). (e) Image histograms of the SEM micrographs in (c) (dashed black curve) and (d) (solid black curve). The less noisy image (d) has  a narrower image histogram.}
    \label{fig:figs1}
\end{figure}

Figure~\ref{fig:figs1}(a) shows an SEM image of spherical tin nanoparticles attached to a copper support structure on a standard TEM grid (image courtesy Navid Abedzadeh). Figure~\ref{fig:figs1}(b) is the histogram of this image. It is a plot of the number of pixels with a certain pixel brightness versus the range of pixel brightnesses. We can see that the pixel brightnesses fall into two broad peaks corresponding to the darker (lower pixel brightness) background pixels and the brighter (higher pixel brightness) pixels representing the tin nanoparticles and copper support structure. Therefore, the histogram reveals the distribution of pixel brightnesses in the image and can be used to quantify noise in it. For example, Figures~\ref{fig:figs1}(c) and~\ref{fig:figs1}(d) are SEM images of the same uniform region of copper, taken at two different pixel dwell times $\tau$. Figure~\ref{fig:figs1}(c) was taken at a small pixel dwell time $\tau=7.5$ \si{\micro\second}, meaning that fewer beam electrons were incident on each pixel of the image, than Figure~\ref{fig:figs1}(d), for which $\tau=28$ \si{\micro\second}. Fewer incident electrons lead to a more noisy signal, and the image in Figure~\ref{fig:figs1}(c) looks noisier than~\ref{fig:figs1}(d). Figure~\ref{fig:figs1}(e), which shows the histogram of these two images, confirms this observation. The image histogram for Figure~\ref{fig:figs1}(c) (dashed curve) is broader than that for Figure~\ref{fig:figs1}(d) (solid curve) indicating that the pixel brightness distribution is narrower for the less noisy image in Figure~\ref{fig:figs1}(d). Note that the means of the two image histograms are the same (pixel brightness level of 90) due to signal time-averaging, as discussed in the paper. 

\section{Optimizing image brightness, contrast, and detector suppressor voltage for in-chamber SE detector}
\label{sec:imageopt}
In the section, we described how we optimized the beam current and pixel dwell time to obtain discrete-SE peaks. Here, we will describe related optimization of the image brightness and contrast settings on the SEM, as well as the in-chamber detector suppressor voltage. We will also describe the effect of beam blanking and turning the beam off on the image histogram.

\begin{figure}[ht]
    \centering
    \includegraphics[scale=0.44]{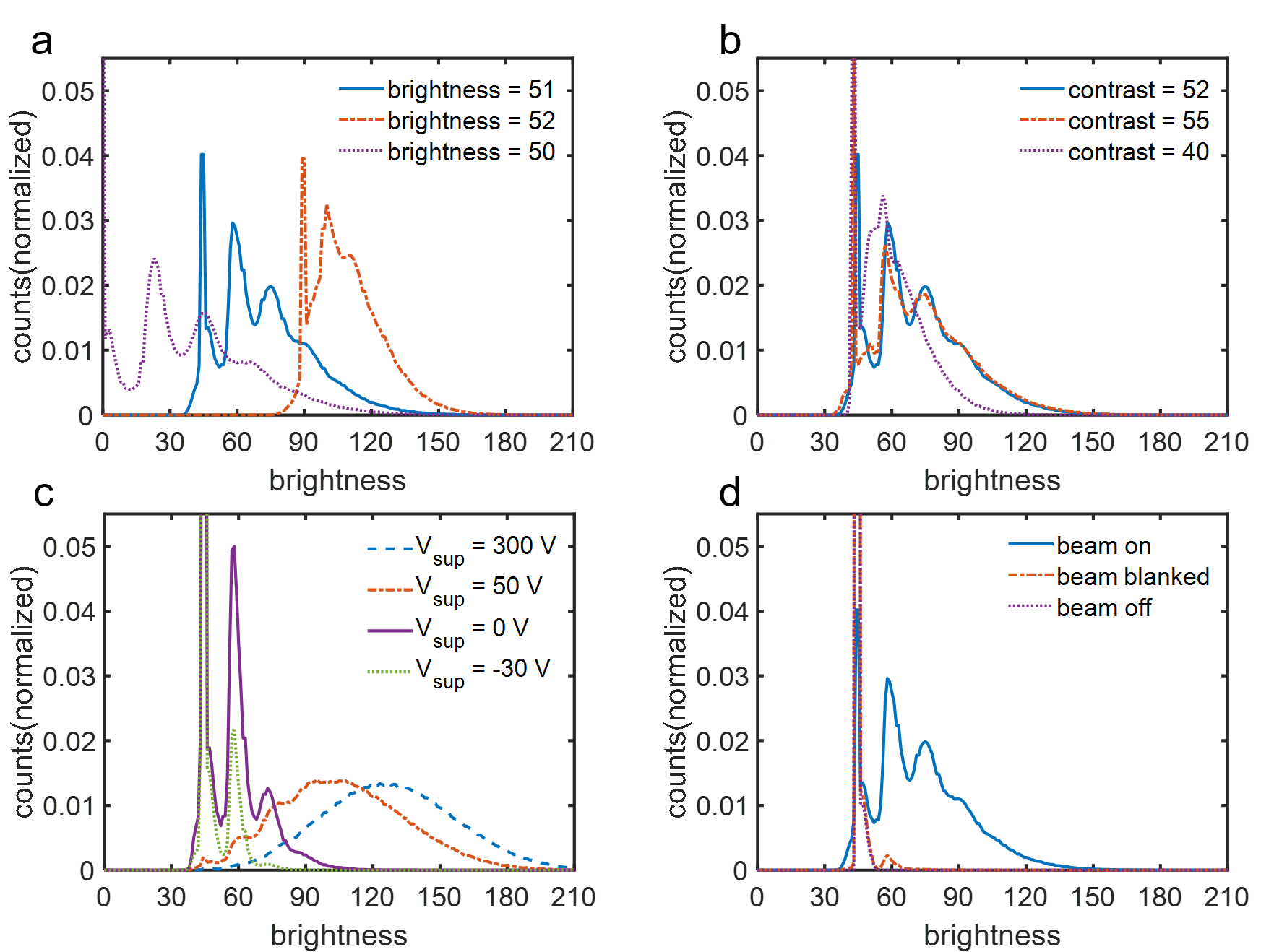}
    \caption[In-chamber SE image histogram optimization]{In-chamber SE image histogram optimization. (a) Optimization of image brightness. An image brightness of 51 (solid blue curve) ensured that the sharp noise peak was well-resolved (unlike image brightness 50, dotted purple curve) and the SE peaks did not merge into one another (unlike brightness 52, dash-dotted orange curve). (b) Optimization of image contrast. A contrast of 52 (solid blue curve) ensured that the peaks did not merge into one another (unlike contrast 40, dash-dotted orange curve) and the noise-artifacts did not become significant (unlike contrast 55, dotted purple curve). (c) Variation of image histogram with in-chamber detector suppressor voltage. As the suppressor voltage was reduced, the SE peaks became more distinct and the mean SE count reduced since fewer SEs were attracted to the detector. (d) Image histograms for beam on (solid blue curve), beam blanked (dash-dotted orange curve), and beam off (dotted purple curve). When the beam is blanked, the SE peaks disappear except for a small residual one-SE peak which disappears when the beam is turned off. The remaining sharp peak corresponds to the zero-SE noise level.}
    \label{fig:figs2}
\end{figure}

\subsection{Image brightness}
\label{subsec:brightnessopt}
The image brightness specifies the DC offset applied on the signal coming from the SE detector. It is different from the pixel brightness which is the 8-bit value of each pixel in the image. The SEM software displays the image brightness as a number between 0 and 100.

Figure~\ref{fig:figs2}(a) shows the change in the image histogram on changing the image brightness on the SEM software. The image brightness values shown are 51 (solid blue curve), 52 (dash-dotted orange curve), and 50 (dotted purple curve). As we would expect from adding a DC offset, the histogram shifted to higher pixel brightness values for higher image brightness, and vice-versa. Further, the peaks in the histogram merged with each other at 52 image brightness. As we will show in Section~\ref{subsec:introosc}, the maximum voltage output from the SE detector at the image contrast settings we used was about \num{5.6} \si{\volt}. Therefore, as we increased the image brightness, the range over which the signal from the detector could vary reduced. The peak merging was a result of this reduced dynamic range of the detector signal. Conversely, at image brightness 50 and below, the sharp noise peak shifts to zero pixel brightness and can no longer be resolved. Based on the results of Figure~\ref{fig:figs2}(c), we decided to use an image brightness of 51 for all experiments to resolve the zero-SE peak and avoid artifacts in our analysis due to some pixels being saturated at zero pixel brightness but not lose the quantization of the peaks due to reduced dynamic range at higher image brightnesses. All of the results reported in the paper were obtained at an image brightness of 51.

\subsection{Image contrast}
\label{subsec:contrastopt}
The image contrast specifies the overall dynamic range over which the signal from the SE detector can vary. On the SEM software, the image contrast is specified as a value between 0 and 100. We obtained the data in Figures~\ref{fig:figs2}(c) at an image contrast of 52.

Figure~\ref{fig:figs2}(b) shows the image histograms for different values of image contrast. These histograms were obtained at an image brightness of 51. The values of image contrast in this plot are 52 (solid blue curve), 40 (dash-dotted orange curve), and 55 (dotted purple curve). We see that reducing the contrast reduced the peak separation and the peaks merged with each other at image contrast of 40. The merging is a result of a lower dynamic range, similar to the effect in our discussion about image brightness. At contrast 55, we started seeing shoulders on both sides of the one-SE peak around pixel brightnesses 50 and 70. We suspect that these shoulders were due to detector noise that gets amplified at high contrast. Based on the results of Figure~\ref{fig:figs2}(b), we decided to use a brightness of 52 for all future experiments to get well-separated SE quantization peaks without introducing artifacts due to detector noise. All of the results reported in the paper were obtained at an image contrast of 52.

\subsection{Detector suppressor voltage}
\label{subsec:suppop}
The suppressor voltage $\mathrm{V_{sup}}$ is applied to the external cage on the in-chamber SE detector to accelerate and attract the low-energy SEs towards it. In Figure~\ref{fig:figs2}(c), we plot the image histograms for $\mathrm{V_{sup}}=300$ V (dashed blue curve), $\mathrm{V_{sup}}=50$ V (dash-dotted orange curve), $\mathrm{V_{sup}}=0$ V (solid purple curve), and $\mathrm{V_{sup}}=-30$ V (dotted purple curve). The change in mean SE count was small when we reduced $\mathrm{V_{sup}}$ from 300 V to 50 V. At $\mathrm{V_{sup}}=300$ V, $N_{SE}=5.59$ and at $\mathrm{V_{sup}}=50$ V, $N_{SE}=4.18$. Most SEs have energies less than 50 eV, so an accelerating voltage above 50 V should be sufficient to attract them to the detector. However, the histogram changed substantially on reducing $\mathrm{V_{sup}}$ from 50 V to 0 V; $N_{SE}=0.78$ at 0 V. Between 50 V and 0 V, the suppressor voltage was in the range of SE energies, and the detector missed more and more of the SEs as the voltage was reduced. At $\mathrm{V_{sup}}=0$ V, there was no voltage to attract the SEs to the detector, and only SEs that were emitted with sufficient energy and in the direction of the detector were detected. Hence, the mean number of SEs went down, and very few pixels registered two or three SEs as seen in the purple curve. On reducing the suppressor voltage to negative values, the electron peaks were further suppressed, and at $\mathrm{V_{sup}}=-30$ V only a small fraction of the pixels registered one SE ($N_{SE}=0.18$). In this condition the detector repelled most SEs and a very small fraction of them made it to the detector. The contribution of $\mathrm{{SE_{III}}}$s was much more significant in this condition due to the relative lack of $\mathrm{{SE_{I}}}$s and $\mathrm{{SE_{II}}}$s at the detector. The sequential suppression of the histogram peaks as the suppressor voltage was reduced is more evidence towards their origin being due to SE quantization.

We used a suppressor voltage of 300 V (the maximum possible value) for the experimental results reported in the paper to maximize the number of SEs that were registered on our detector.

\subsection{Beam blanking}
\label{subsec:blankingop}
In Figure~\ref{fig:figs2}(d), we plot the image histogram in three cases: beam on (solid blue curve), beam blanked (dotted orange curve), and beam off (dotted purple curve). We see that when the beam was blanked, most of the pixels had a brightness of 44 except for a small peak at the one SE brightness level of 58. When the incident beam was switched off this small peak disappeared. When the incident beam is blanked in the SEM, deflection plates near the electron gun prevent it from reaching the sample; however, the beam is still on. Therefore, it is possible for stray SEs from higher up in the column to make it to the detector leading to the small peak. Note the similarity between this histogram and the histogram obtained at $\mathrm{V_{sup}}=-30$ V in Figure~\ref{fig:figs2}(c). In both cases, a few SEs made it onto the detector leading to the small one-SE peak. When the beam was switched off, no secondaries were generated anywhere in the column which led to the one-SE peak disappearing. The only peak now was at pixel brightness 44. From this observation, we concluded that this peak was due to detector noise and corresponds to the zero-SE level.

\section{Joy's method of calculating image SNR and DQE in SEM}
\label{sec:joymethod}

\begin{figure}
    \centering
    \includegraphics[scale=0.44]{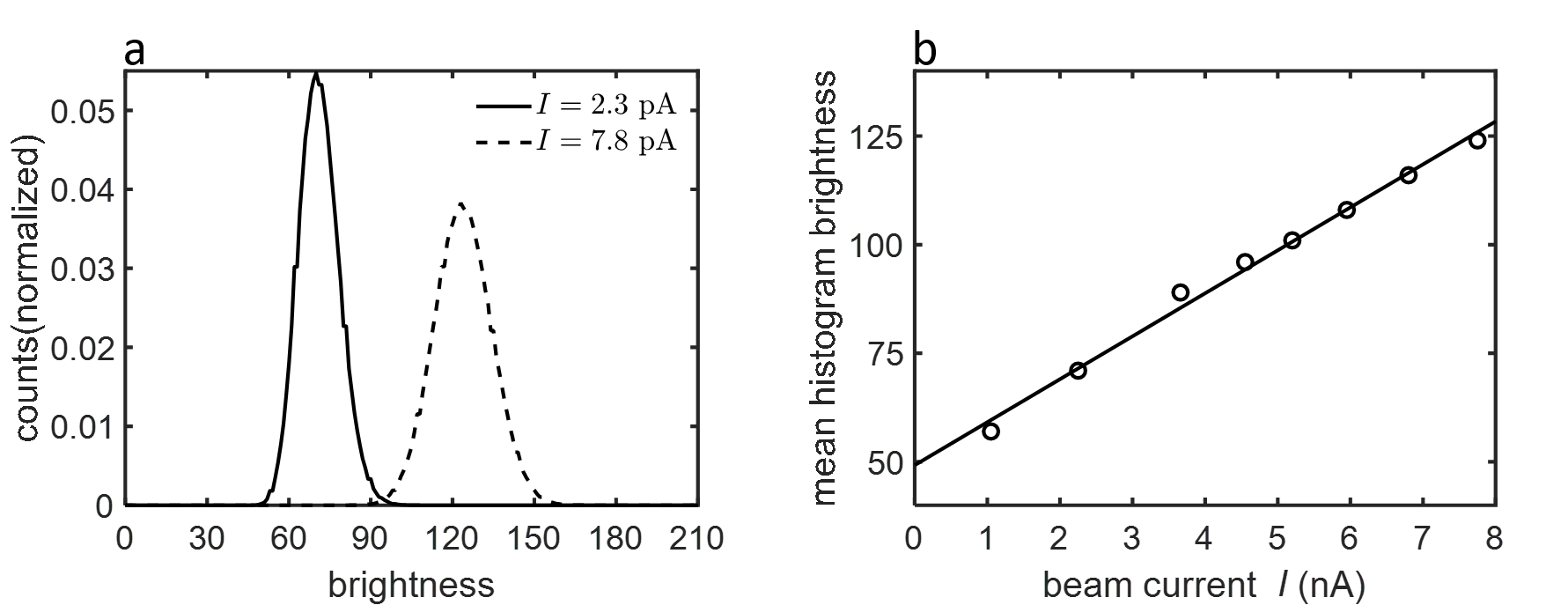}
    \caption[Joy's method of calculating image SNR and SE detector DQE]{Joy's method of calculating image SNR and SE detector DQE. (a) Histograms of two SEM images of the same uniform aluminum sample taken at two incident beam currents: $2.3$ \si{\pico\ampere} and $7.8$ \si{\pico\ampere}. The higher current histogram has a higher mean pixel brightness as well as higher width. (b) Variation of mean histogram brightness with incident beam current. The vertical intercept of the least-square fit line is the offset level due to the image brightness and contrast settings.}
    \label{fig:figs3}
\end{figure}

Joy's method of finding the DQE of SE detectors in SEM relies on using image histograms to calculate the SNR at different imaging currents. Figure~\ref{fig:figs3}(a) shows image histograms for the in-chamber detector from a uniform region of aluminum foil obtained at an incident beam energy of 10 kV and $\tau=28$ \si{\micro\second}, a working distance of 13 mm, at two beam currents: $I=2.3$ \si{\pico\ampere} (solid black curve) and $I=7.8$ \si{\pico\ampere} (dashed black curve). The mean pixel brightness for the low-current histogram for is 71 and for the high-current histogram is 124. We expect the mean pixel brightness to be higher for the image histogram at higher incident beam current due to higher $N_{SE,\textrm{object}}$. Further, the histogram for $I=2.3$ \si{\pico\ampere} is narrower than the histogram for $I=7.8$ \si{\pico\ampere}; the full-width-at-half maximum (FWHM) for the $I=2.3$ \si{\pico\ampere} histogram is 18 pixel brightness units, while the FWHM for the $I=7.8$ \si{\pico\ampere} histogram is 24 pixel brightness units. For a Poisson-distributed random variable, the variance is equal to the mean. Hence, the width of its probability distribution increases as the square root of the mean. Therefore, as $N_{SE,\textrm{object}}$ increases due to higher current, we expect the histogram mean to increase linearly and the histogram FWHM to increase as the square root of the current. For an ideal detector (DQE $=1$), we would expect the SNR to be equal to $\sqrt{N_{SE,\textrm{object}}}$. We will refer to this quantity as $\textrm{SNR}_{\textrm{ideal}}$.

Before we can use our image histograms to calculate SNR, we have to correct the histograms for the image brightness and contrast settings we used. We would expect the image histogram mean to go to zero as the beam current $I$ reduces to zero. However, the specific image brightness and contrast settings we use might offset this value and make it non-zero. In order to find what this offset level is, we plot the image histogram mean as a function of the incident beam current in Figure~\ref{fig:figs3}(c). As expected, the variation of the image histogram mean with $I$ was linear. We extrapolated the least-square fit line (shown in black) to zero current to find the offset level. In this case, the offset level was 49 pixel brightness units. This value is close to the noise level (44) for the in-chamber detector at these settings, as reported in the paper. Next, we subtracted the offset level from the image histogram means to find the true means at zero offset. The ratio of the corrected histogram mean to the FWHM gave us $\textrm{SNR}_{\textrm{exp}}$, which is the experimental SNR. $\textrm{SNR}_{\textrm{exp}}$ includes the effect of the non-unity DQE and is equal to $\sqrt{N_{SE}}$, the SNR for a Poisson process with mean equal to the number of SEs that are detected at the in-chamber detector. From Joy's work,
\begin{equation*}
    N_{SE}=N_{SE,\textrm{object}}\cdot\textrm{DQE}.
\end{equation*}
using our definitions of  $\textrm{SNR}_{\textrm{ideal}}$ and $\textrm{SNR}_{\textrm{exp}}$, we get
\begin{equation*}
    \textrm{SNR}_{\textrm{exp}}^2=\textrm{SNR}_{\textrm{ideal}}^2\cdot\textrm{DQE}.
\end{equation*}
Hence,
\begin{equation*}
    \textrm{DQE}=\textrm{SNR}_{\textrm{exp}}^2/\textrm{SNR}_{\textrm{ideal}}^2.
\end{equation*}
From the image histograms for the different incident beam currents shown in Figure~\ref{fig:figs3} we extracted a DQE between 0.15 and 0.22 depending on the incident beam current. We ascribe this variation to small non-linearities in the detector at higher currents as discussed further in Section~\ref{subsec:oschisto}. We also performed a similar analysis for the in-lens detector and obtained a DQE between 0.4 and 0.6. In these calculations, we used $\delta=0.2$ for our aluminum sample. The values of DQE extracted from our implementation of Joy's method are in the range reported previously for well-designed in-chamber and in-lens detectors and serve as a benchmark for our technique of calculating DQE in the paper.

\section{In-lens detector statistics}
\label{sec:inlens}
In this section, we will show the variation in the image histogram for the in-lens detector with changing imaging conditions. We presented similar data for the in-chamber detector in the paper. This data shows evidence of integral-SE-count peaks in the image histograms for the in-lens detector and supports our conclusions in the paper.

\begin{figure}
    \centering
    \includegraphics[scale=0.44]{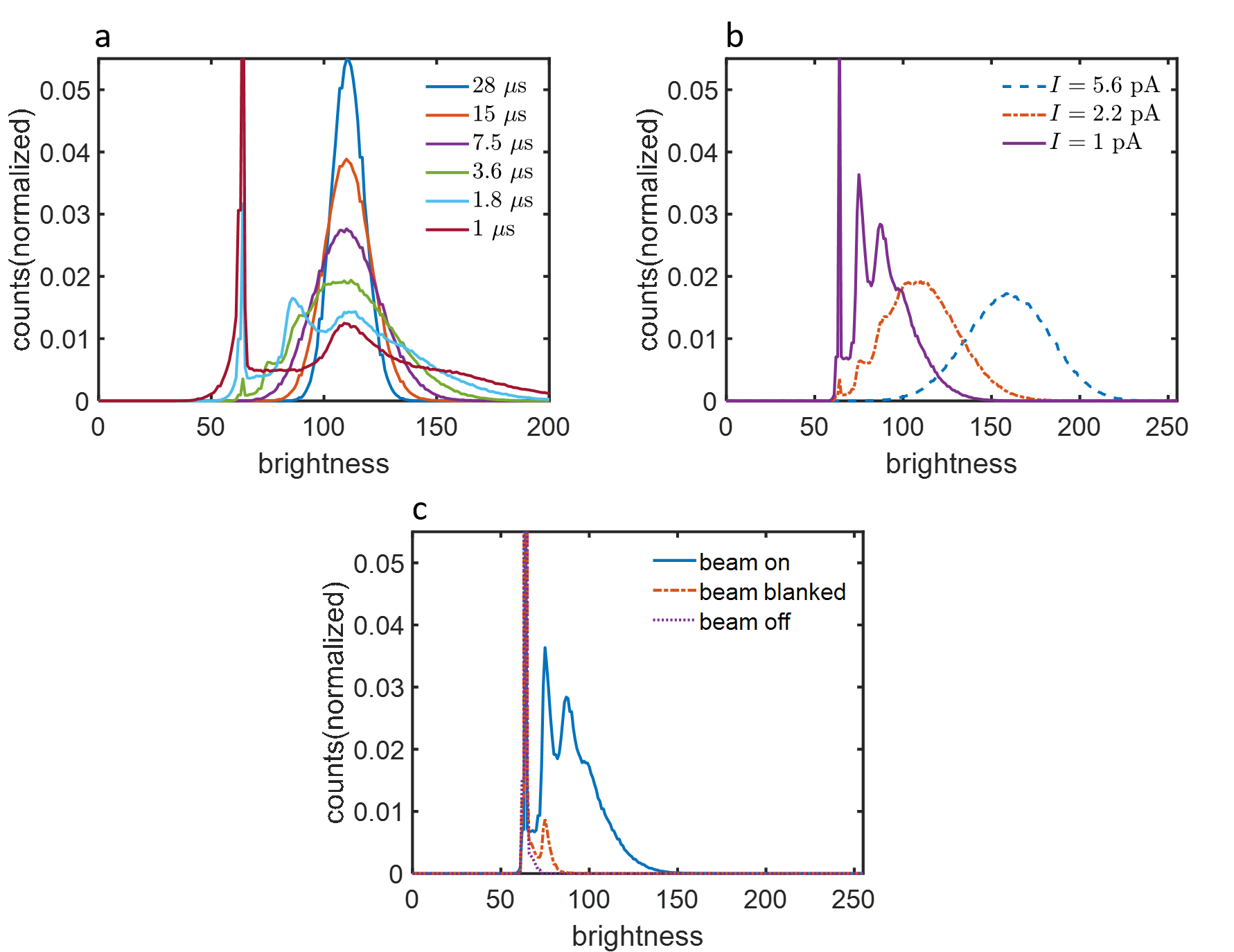}
    \caption[In-lens image histogram]{In-lens image histogram. (a) Variation in image histogram with changing pixel dwell time. As the pixel dwell time was lowered from 28 \us~to 1 \us, distinct peaks appeared in the histogram corresponding to 0, 1 and 2 SEs. (b) Variation in image histogram with changing incident electron beam current at a pixel dwell time of 3.6 \us. Discrete peaks appear in the histogram at 1 \pA beam current corresponding to integral number of SEs. (c) Variation in the image histogram when the beam is on, blanked and off. }
    \label{fig:figs4}
\end{figure}

Figure~\ref{fig:figs4}(a) shows the variation in the in-lens detector image histogram with pixel dwell time. We obtained these histograms by scanning the electron beam over a uniform sample of aluminum at a beam current of 2.2 \pA~, beam energy of 10 \keV, and working distance of 13 mm. Just as for the in-chamber detector, at high pixel dwell time the image histogram is nearly Gaussian. As the pixel dwell time reduces from 28 \us~to 7.5 \us, the mean of the histogram remains at 111 but the width increases. At a pixel dwell time of 3.6 \us, we see discrete peaks begin to emerge; there is a sharp peak at brightness 64 and a broader one at brightness 75 in the histogram. The histogram for 1.8 \us~shows a sharp peak at 64, and broad peaks at 87 and 112. Finally, the 1.8 \us~histogram shows a sharp peak at 64 and a broad peak at 112. The gap between consecutive peaks is 11 pixel brightness units for the 3.6 \us~histogram, $\sim24$ for the 1.8 \us~histogram, and 48 for the 1 \us~histogram. Just as we had observed for the in-chamber image histograms, the gap between peaks doubled when the pixel dwell time was halved due to signal time-averaging. Therefore, we can ascribe these peaks to integral number of SEs, with the sharp peak corresponding to 0 SEs.

Figure~\ref{fig:figs4}(b) shows the change in the in-lens image histogram with incident beam current. Each of these histograms was taken at a pixel dwell time of 3.6 \us. As the incident beam current reduces from 5.6 \pA~to 1 \pA, peaks corresponding to integral number of SEs appear in the histogram.

Figure~\ref{fig:figs4}(c) shows the in-lens histogram when the beam is on, blanked and off. Just as for the in-chamber image, a few pixels register one SE when the beam is blanked due to SEs generated in the electron beam column. When the beam is switched off, these SEs disappear and only the sharp, 0 SE peak remains. This result confirms that the sharp peak corresponds to noise and dark counts in the detector.

\begin{figure}
    \centering
    \includegraphics[scale=0.44]{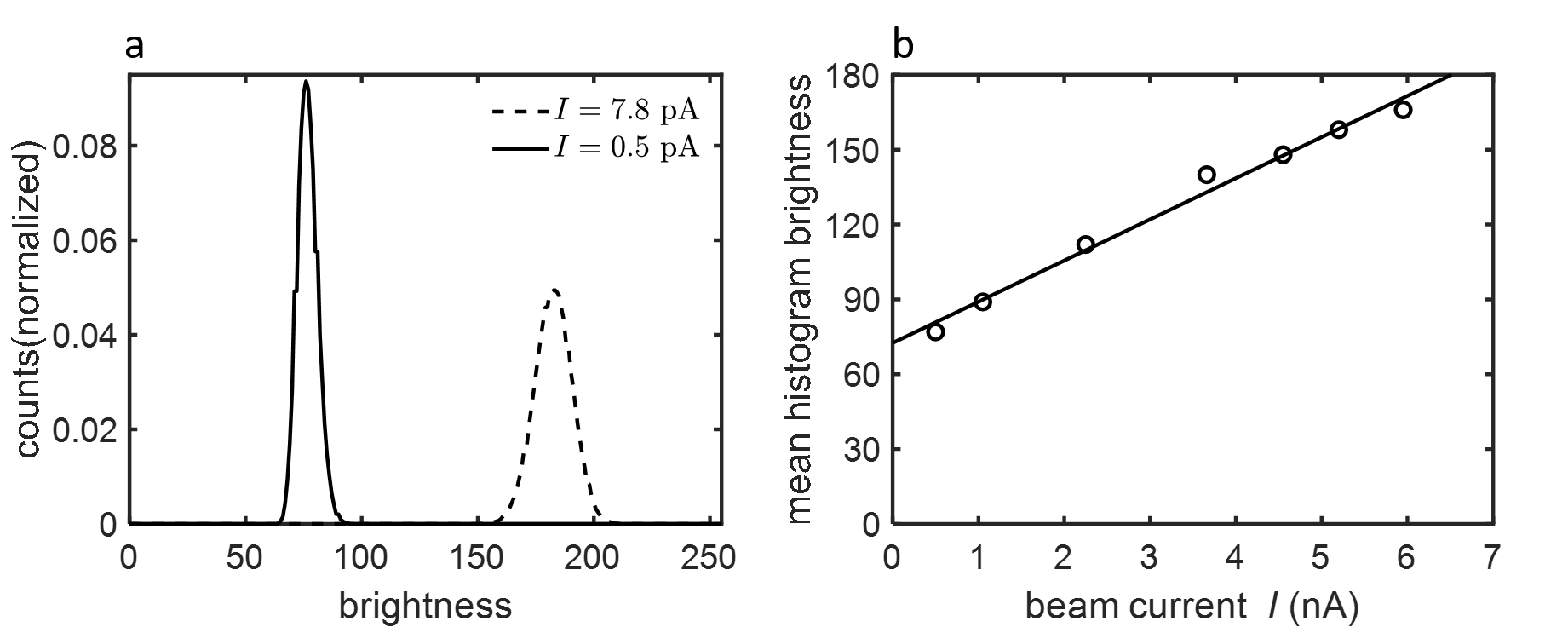}
    \caption[Joy's method for in-lens detector]{Joy's method of finding DQE for the in-lens detector. (a) Change in image histogram on changing the electron beam current from 7.8 \pA~to 0.5 \pA for a pixel dwell time of 28 \us. The histogram mean lowers from 183 to 76 and the histogram gets narrower. (b) Variation in mean histogram brightness as a function of beam current.}
    \label{fig:figs5}
\end{figure}

In Figure~\ref{fig:figs5}, we show the implementation of Joy's method for finding DQE for the in-chamber detector. The images corresponding the these histograms were acquired for a pixel dwell time of 28 \us, with varying incident beam current. Figure~\ref{fig:figs5}(a) shows the image histograms for two values of the beam current. The histogram for $I=7.8$ \pA~has a mean of 183, while that for $I=0.5$ \pA~has a mean of 76. From these histogram mean values, we extracted the mean SE counts just as for the in-chamber detector. Figure~\ref{fig:figs5}(b) is a plot of the extracted mean values (unfilled black circles) as a function of the incident beam current. The solid black line is a least-squares fit line and gives us the offset in mean values due to the contrast and image brightness values we used. This value is 72.6, close to the 0 SE-level of 64 we found from the histograms in Figure~\ref{fig:figs4}. We used the offset-corrected mean pixel brightness values to find the in-lens DQE reported in the paper.

\section{SE counting using oscilloscope outcoupling}
\label{sec:oscilloscope}
In the paper we described our scheme and observations of discrete SE peaks using the image histogram. We presented evidence for the observed histogram peaks arising due to SEs and applied these findings to calculate the DQE of the in-chamber and in-lens SE detectors and map the variation of the DQE with working distance. In this section, we will provide further experimental evidence supporting our observation of discrete SE peaks in the image histogram technique by directly analyzing the signal from the SE detectors by outcoupling it onto an oscilloscope. We performed this outcoupling to reproduce the image histograms using the direct time-domain signal from the SE detector and rule out the possibility of the histogram peaks arising from an artifact of the SEM imaging algorithm. In addition to providing further evidence of SE quantization in the image histogram, this technique suggests further extensions of this work towards live electron count imaging. As we had done in the discussion of the image histogram technique, we will focus on the in-chamber SE detector here; we obtained similar results for the in-lens detector.

\subsection{Introduction to oscilloscope outcoupling}
\label{subsec:introosc}

\begin{figure}
    \centering
    \includegraphics[scale=0.44]{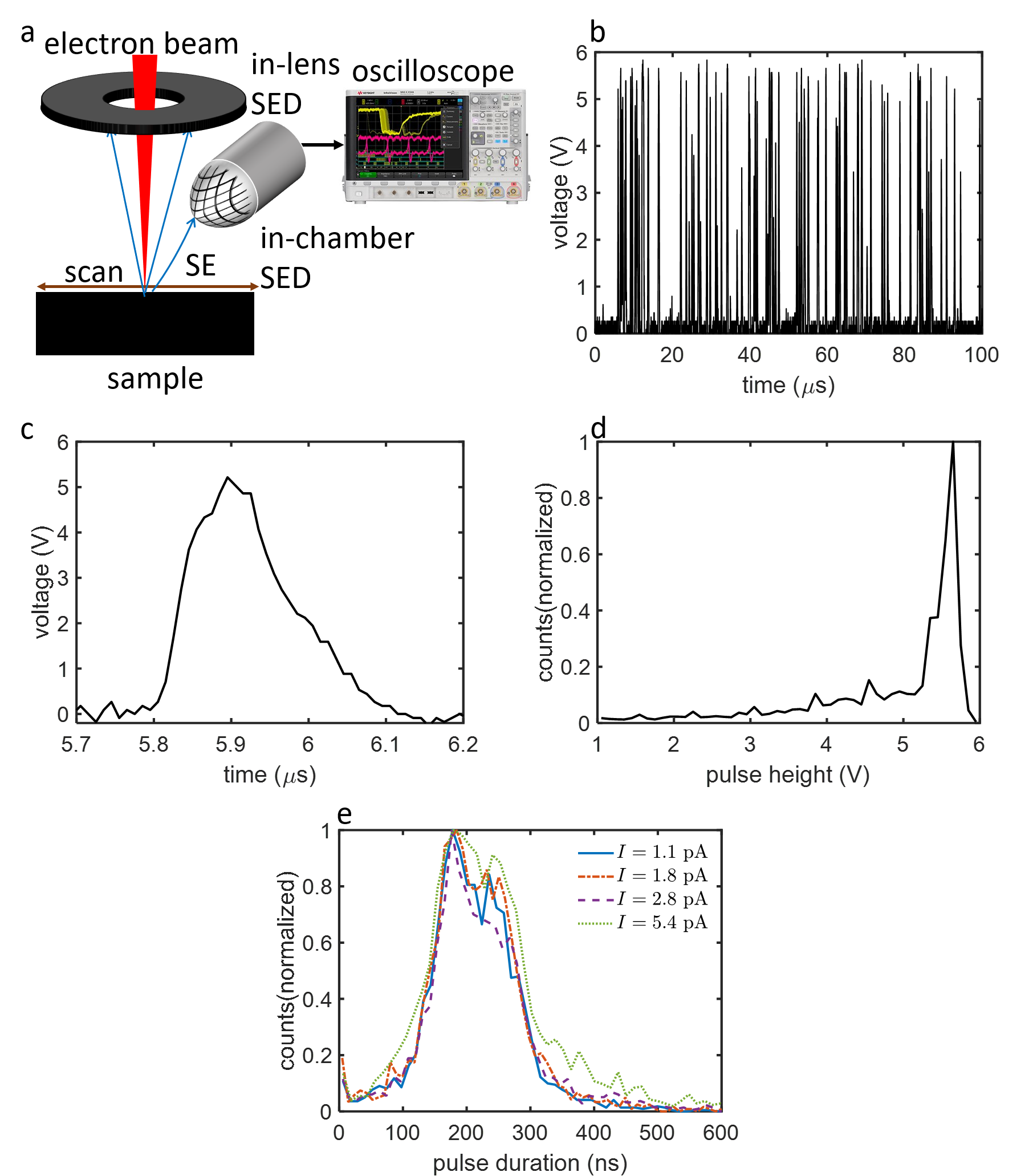}
    \caption[Oscilloscope outcoupling of SE detector signal]{Oscilloscope outcoupling of SE detector signal. (a) Scheme for oscilloscope outcoupling. The signal from the SE detector is directly analyzed on an oscilloscope. (b) $100$ \si{\micro\second} output signal from the in-chamber detector on the oscilloscope (sampled at 10 \ns) showing pulses due to detected SEs. (c) A single signal pulse with a FWHM time duration of $148$ \ns. (d) Histogram of signal pulse heights showing that most pulses are saturated at 5.6 V. (e) Histograms of the FWHM pulse durations for $I$ between $1.1$ \si{\pico\ampere} and $5.4$ \pA. All histograms have a mean of $180$ \ns.}
    \label{fig:figs6}
\end{figure}
Figure~\ref{fig:figs6}(a) is a schematic of the oscilloscope outcoupling scheme we used. In this experiment, we held our beam stationary over one spot on a uniform sample of aluminum. Since the sample is uniform and the beam is held stationary, we did not need to synchronize the collection of detector signals on the oscilloscope with the SEM scanning; we assumed that the expected number of SEs for a time window corresponding to the pixel dwell time did not vary over the collection period. The SE signal from the in-chamber detector was coupled to a 2 GHz LeCroy WaveRunner 6200A oscilloscope through a standard BNC cable. As described in more detail in Section~\ref{subsec:oschisto}, we used the outcoupled signal to generate histograms on the oscilloscope and compared these histograms with the image histograms generated from the SEM image.  

The outcoupling port on our detector was present after a pre-amplification stage which meant that the signal we obtained on the oscilloscope was not the raw signal from the photomultiplier tube. 
As we will describe in more detail later, we observed one pulse per incident SE with a mean time duration of $\approx 180$ \si{\nano\second}. These long pulses were a result of the low-pass filtering and amplification applied at the pre-amplification stage. Therefore, we did not observe pulses from individual photoelectrons in the photomultiplier-scintillator setup in the detector, excited by an SE. However, for the purpose of our experiment, the longer-time pulse generated by each detected SE was sufficient as long as pulses from successive SEs did not overlap with each other, which was true at low beam currents.

\begin{figure}[ht]
    \centering
    \includegraphics[scale=0.44]{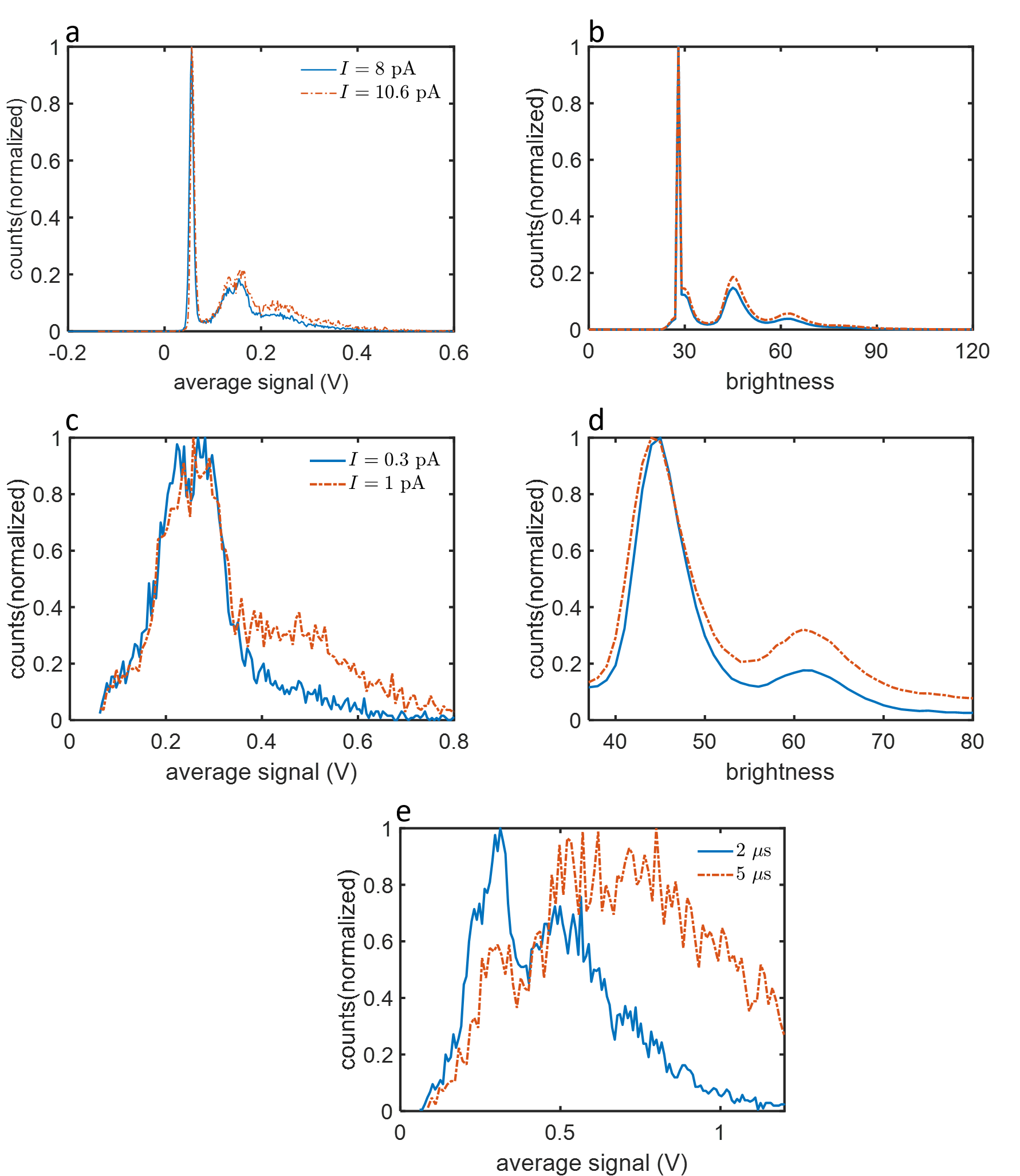}
    \caption[Oscilloscope histograms and their statistics.]{Oscilloscope histograms and their statistics. (a) Oscilloscope histogram of the average signal level for $I= 8$ \si{\pico\ampere} (solid blue curve) and $I= 10.8$ \si{\pico\ampere} (dash-dotted orange curve). Both histograms show a sharp noise peak and two broad integral SE peaks with the intensity of the SE peaks being higher for the higher incident beam current histogram. The mean SE number for the $I= 8$ \si{\pico\ampere} histogram was 0.94 and for the $I= 10.8$ \si{\pico\ampere} it was 1.24. (b) Image histograms for the same object region at the same incident beam currents as (a). The mean SE number for the $I= 8$ \si{\pico\ampere} histogram was 1.07 and for the $I= 10.8$ \si{\pico\ampere} it was 1.25. (c) Oscilloscope histogram acquired with an oscilloscope trigger voltage of 0.6 V, for $I= 0.3$ \si{\pico\ampere} (solid blue curve) and $I= 1$ \si{\pico\ampere} (dash-dotted orange curve). The higher current histogram shows higher counts in the two-SE peak region.(d) Image histograms for the same object scan regions for the same incident beam currents as (c). The ratio of the one- and two-SE peaks is $\sim5$ at $I= 0.3$ \si{\pico\ampere} and $\sim2.5$ at $I= 1$ \si{\pico\ampere} for both the oscilloscope and image histograms. (e) Oscilloscope histograms for $5\times10^3$ detector signal windows of duration 2 \si{\micro\second} (solid blue curve) and 5 \si{\micro\second} (dash-dotted orange curve) with trigger voltage of 0.6 V. Both histograms show distinct SE peaks, and the higher duration histogram has higher intensity for two-SEs and higher SE numbers indicating a higher mean SE count.}
    \label{fig:figs7}
\end{figure}

From Figure~\ref{fig:figs6}(b) we can also see that the pulses had varying time durations. In Figure~\ref{fig:figs6}(e), we plot histograms of the pulse time duration of $2\times10^4$ observed pulses, with the time duration measured at the pulse FWHM, for incident beam currents between 1.1 \si{\pico\ampere} and  5.4 \si{\pico\ampere}. The distribution of pulse time duration was approximately Gaussian with a mean duration of 180 \si{\nano\second}. Moreover, the distribution remained the same for different incident beam currents up to 8 \si{\pico\ampere} (not plotted here). As the current increased, the average number of pulses per time window increased, but their width followed the same distribution as in Figure~\ref{fig:figs6}(e). Therefore, we concluded that each pulse was the result of a single SE detection, and the number of pulses in each time window corresponded to the number of SEs incident on the detector in that time.

As described in the paper, the maximum beam current we used in our experiments was about 8 \si{\pico\ampere} and $\tau=3.6$ \si{\micro\second}. Assuming that $\delta=0.2$ and the DQE for the in-chamber detector is 0.16, $N_{SE}=5.8$. Therefore, there is a high probability of multiple SEs incident within the same pixel dwell time. The voltage pulses corresponding to these SEs may overlap enough to cause a significant non-linear distortion in the output waveform. We can ascribe the deviation from linearity of our mean SE count measurements in the paper to the increasing probability of pulse overlap at higher incident beam currents.

\subsection{Oscilloscope histograms and statistics}
\label{subsec:oschisto}

Once we had determined the origin of the pulses in the outcoupled signal on the oscilloscope, we could start comparing its statistics to those from the image histogram. In an SEM image, the brightness of each pixel on the SEM image corresponds to the average signal level from that pixel. Therefore, we collected a series of $10^4$ signal windows on the oscilloscope, each of time duration $5$ \si{\micro\second}, sampled at $10$ \si{\nano\second}. The total collection time for all the signal windows was about 40 minutes. We were limited in the number of signal windows we could collect by the memory of the oscilloscope and contamination buildup as discussed later in this section. We used random triggering on the oscilloscope to ensure that the collected signal was not biased. However, a small fraction ($<2\%$) of the signal windows had incomplete SE pulses at either the beginning or end of the window collection time. We did not consider these windows in our analysis of oscilloscope histograms. Since the fraction of such windows was small, we do not expect their omission to impact the statistics we will report here.

In Figure~\ref{fig:figs7}(a), we plot the histogram of the average signal level from the collected signal windows. Note that this histogram (and others in this figure) is normalized to the height of the highest peak instead of the histogram area which was the case for the histograms we have discussed in the paper. We made this change to make it easier to visualize the differences between the histograms at different incident beam currents. The solid blue curve in Figure~\ref{fig:figs7}(a) corresponds to a beam current of 8 \si{\pico\ampere} and the dash-dotted orange curve corresponds to 10.8 \si{\pico\ampere}. The histograms for both currents exhibited a sharp peak at an average voltage value of 0.06 and two broad peaks at average voltages of 0.15 and 0.24 V. For comparison, in Figure~\ref{fig:figs7}(b) we plot image histograms for the same currents with the same settings. Note that the image brightness in this histogram was slightly different than in the other figures in this paper (50.9 instead of 51) leading to different peak positions. The image histograms show the usual sharp noise peak and distinct one- and two-SE peaks. The presence of these peaks in the histogram generated from the oscilloscope signal, as well as the similarity between the overall shape and regular spacing of the peaks in the two types of histograms, provides further evidence that the origin of these peaks is SE quantization. We also note that the oscilloscope histogram appears to be noisier because of the lower number of signal windows used to generate it ($10^4$) compared to the SEM images which are $1024\times768$ pixels.

We can use the oscilloscope histograms to estimate the mean number of SEs using the same procedure we had discussed for image histograms in the paper. Using this process, the mean SE count for the 8 \si{\pico\ampere} oscilloscope histogram was 0.94. The mean SE count for the 10.8 \si{\pico\ampere} oscilloscope histogram was 1.24. In comparison, the mean SE count for the 8 \si{\pico\ampere} image histogram was 1.07, and the mean SE count for the 10.8 \si{\pico\ampere} image histogram was 1.25. Apart from fluctuations due to the lower number of signal windows used to generate the oscilloscope histogram, the deviations between the oscilloscope and image histogram SE counts were caused by two additional reasons. First, there was significant sample contamination buildup during the long (40 minute) data collection time. We noticed a significant change in the mean SE count from image histograms taken at the start of the data collection and the end. For example, for the  10.8 \si{\pico\ampere} dataset, the mean SE count changed from 1.43 at the start of the data collection to 0.86 by the end. The image histograms in Figure~\ref{fig:figs7}(b) are averaged between the two histograms taken at the start and end of the data collection as are the mean SE counts reported from these averaged image histograms. Since the statistics for the oscilloscope histograms in Figure~\ref{fig:figs7}(a) build up over the data collection time, we compared the mean SE number from these histograms to the averaged image histograms. The second reason for deviations between the oscilloscope and image histogram mean values is signal pileup, \textit{i.e.}, the non-linearity due to overlapping pulses at high incident beam current. At beam currents of 8 \si{\pico\ampere} and 10.8 \si{\pico\ampere} the non-linearity can be significant leading to distortion in the oscilloscope and image histograms and their statistics. 

Figure~\ref{fig:figs7}(c) shows oscilloscope histograms of $5\times10^3$ signal windows of duration 5 \si{\micro\second} taken at lower beam currents to mitigate the effect of contamination buildup and detector non-linearity. The solid blue curve is for a beam current of 0.3 \si{\pico\ampere}, and the dash-dotted orange curve is for a beam current of 1 \si{\pico\ampere}. Further, to collect this data we used a trigger voltage of 0.6 V on the oscilloscope (instead of using random triggering as in Figure~\ref{fig:figs7}(a)) to filter out the noise pulses and get better statistics on the higher SE number pulses. The oscilloscope for $I=1$ \si{\pico\ampere} shows a higher signal in the two-electron peak region. Due to the lack of pulses lower than 0.6 V we could not obtain a mean SE count from these histograms. However, we can compare the ratio of the one- and two-SE peaks from these oscilloscope histograms to the values from the image histograms. Figure~\ref{fig:figs7}(c) is an image histogram for the same scan region under the same settings. The ratio of the one- and two-SE peak for both the oscilloscope and image histograms at 0.3 \si{\pico\ampere} was approximately 5, and the ratio for the 1 \si{\pico\ampere} histograms was approximately 2.5. Although the SE peaks in the oscilloscope histogram are not as well resolved as in the image histogram due to fewer samples, this equality between the one- and two-electron peak ratios indicates that the statistics from both types of histograms are equivalent.  

In Figure~\ref{fig:figs7}(e), we plot the oscilloscope histogram for $5\times10^3$ pulses for sampling window times of 2 \si{\micro\second} (solid blue curve) and 5 \si{\micro\second} (dash-dotted red curve). We obtained both histograms with the oscilloscope trigger voltage set to 0.6 V for an incident beam current of 4 \si{\pico\ampere}. Note that we used a larger aperture on the SEM (20 \si{\micro\meter} instead of 7.5 \si{\micro\meter} for all earlier results) to get higher SE signal levels. The 2 \si{\micro\second} histogram shows two clear peaks at average signal levels 0.27 V and 0.51 V. The  5 \si{\micro\second} histogram also shows peaks at these values and higher signal at 0.51 V and above. Again, these results are consistent with our interpretation of the peaks as arising from single SEs. The reduction in histogram counts in going from 5 \si{\micro\second} to 2 \si{\micro\second} is similar to the emergence of SE peaks in Figure~\ref{fig:figs2}(a) showing the consistency between the two techniques.
